\documentclass[12pt]{article}
\usepackage{jheppub}
\pdfoutput=1
\usepackage{epstopdf}

\usepackage{mathrsfs}
\usepackage{psfrag}
\usepackage{color}
\usepackage{hyperref}

\usepackage{slashed}
\usepackage{feynmp-auto}
\usepackage{simplewick}
\usepackage{cancel}

\usepackage[table]{xcolor}

\usepackage{amsmath,bbm,array,amsfonts,graphicx,wrapfig,arydshln,lscape,float,multirow,longtable,rotating,makecell}
\usepackage{url}

\let\ifpdf\relax
\usepackage[T1]{fontenc}
\usepackage[utf8]{inputenc}
\usepackage{ifpdf,tocloft,rotating}
\usepackage{hyperref}
\let\normalcolor\relax
\hypersetup{pdftitle={},pdfcreator={},linkcolor=[rgb]{0.15,0.35,0.75},colorlinks=true,citecolor=[rgb]{0.675,0,0.2},urlcolor=[rgb]{0.15,0.35,0.65}}
\let\olditemize\itemize\renewcommand{\itemize}{\vspace{-2pt}\olditemize\setlength{\itemsep}{1pt}\setlength{\parskip}{0pt}\setlength{\parsep}{-0pt}}
\let\oldenumerate\enumerate\renewcommand{\enumerate}{\vspace{-4pt}\oldenumerate\setlength{\itemsep}{1pt}\setlength{\parskip}{0pt}\setlength{\parsep}{0pt}}

\renewcommand{\bar}{\overline}
\renewcommand{\hat}{\widehat}
\renewcommand{\tilde}{\widetilde}

\newcommand{\dlog}{ d \log}

\definecolor{mhvBlue}{rgb}{0.3,0.2,0.75}
\definecolor{fRed}{rgb}{0.48,0.02824,0.18824}
\definecolor{cut2}{rgb}{0.18824,0.18824,0.48}
\definecolor{cut1}{rgb}{0.48,0.02824,0.18824}

\newcommand{\N}{\mathcal{N}}

\renewcommand{\O}{\mathcal{O}}

\newcommand{\J}{\mathcal{J}}

\def\det{\mathop{\rm det}}

\def\eps{\epsilon}

\def\half{\frac{1}{2}}
\usepackage{mathtools}

\def\zb{\bar z}

\title{Logarithmic forms and differential equations for Feynman integrals}
\author[1]{Enrico Herrmann,}
\affiliation[1]{ SLAC National Accelerator Laboratory, Stanford University, Stanford, CA 94039, USA}
\emailAdd{eh10@stanford.edu}

\author[2]{Julio~Parra-Martinez}
\affiliation[2]{Mani L. Bhaumik Institute for Theoretical Physics,\\
UCLA Department of Physics and Astronomy, Los Angeles, CA 90095, USA}
\emailAdd{jparra@physics.ucla.edu}

\abstract{
  We describe in detail how a $\dlog$ representation of Feynman integrals leads to simple differential equations. We derive these differential equations directly in loop momentum or embedding space making use of a localization trick and generalized unitarity. For the examples we study, the alphabet of the differential equation is related to special points in kinematic space, described by certain cut equations which encode the geometry of the Feynman integral. At one loop, we reproduce the motivic formulae described by Goncharov \cite{Goncharov:1996tate} that reappeared in the context of Feynman integrals in \cite{Spradlin:2011wp,Arkani-Hamed:2017ahv,Abreu:2017enx}. The $\dlog$ representation allows us to generalize the differential equations to higher loops and motivates the study of certain mixed-dimension integrals. 
}

\preprint{\begin{flushright} \end{flushright}}

\begin{document}

\maketitle

%
\section{Introduction}
%
%
Scattering amplitudes are objects of central interest in high energy physics and offer crucial insights into the inner workings of quantum field theory (QFT) itself. Besides theoretical explorations into a reformulation of perturbation theory (see e.g.~\cite{ArkaniHamed:2012nw,Arkani-Hamed:2013jha}), scattering amplitudes are relevant for precision collider physics. Describing the interactions of elementary particles at hadron colliders such as the LHC with high accuracy requires the calculation of perturbative corrections to physical observables. In the context of quantum field theory, these perturbative corrections involve the evaluation of Feynman loop integrals which have been the subject of enormous interest since the early days of QFT.

Traditionally, there are a number of techniques available on the market to deal with the evaluation of Feynman integrals. From numeric approaches (see e.g.~\cite{Passarino:2001wv}) to a wide variety of analytic methods (for a summary, see~\cite{Smirnov:2012gma}), all of these tools have their strengths and weaknesses. Having a diverse set of techniques available is often crucial to successfully deal with a given problem at hand. In this work, we would like to add one new item to that list. The differential equations we are going to describe here, are based on the $\dlog$-integrand representation that is available for certain Feynman integrals (for examples, see \cite{ArkaniHamed:2012nw,Bern:2014kca}). 

In the past, $\dlog$-forms have played an important role \cite{ArkaniHamed:2012nw,Arkani-Hamed:2014via,Bern:2014kca} for maximally helicity violating (MHV) amplitudes in planar maximally supersymmetric Yang-Mills theory ($\N=4$ sYM) and are related to a beautiful geometric approach to scattering amplitudes in terms of Grassmannian geometry \cite{ArkaniHamed:2012nw} and the amplituhedron \cite{Arkani-Hamed:2013jha}. Furthermore, it was observed that, in certain cases, the $\dlog$-structure of loop integrands survives beyond the planar limit \cite{Arkani-Hamed:2014via,Bern:2014kca,Bern:2015ple} lending support to the conjectures of nonplanar analogs of dual conformal symmetry as well as a geometric amplituhedron-like picture. 

Here, we go beyond properties of scattering amplitudes and study \emph{$\dlog$ integrals}, $I =\int \Omega$. These integrals are associated with a differential integrand form, $\Omega$, that only has \emph{logarithmic singularities},
\begin{align}
\label{eq:dlog_general}
\Omega(x_1,...,x_n) \to \frac{dx_i}{x_i-a}\, \widetilde\Omega (x_1,...,\hat x_i,...,x_n)\,,
\end{align}
near any pole $x_i\to a$, where the $(n-1)$-form $\widetilde\Omega (x_1,...,\hat x_i,...,x_n)$ is independent of $x_i$.  

Within the method of differential equations \cite{Kotikov:1990kg,Remiddi:1997ny,Gehrmann:1999as}, it is often possible to choose a basis of master integrals where the dependence on the dimensional regularization parameter factorizes \cite{Henn} (differential equations in canonical form). In a number of highly nontrivial examples, it was shown that a good choice of master integrals is related to the existence of logarithmic singularities at the integrand level which can be checked algorithmically, see e.g.~\cite{Henn:2014qga,Wasser:2016}. Given an $n$-form 
\begin{align}
\label{eq:def_rat_form}
\Omega = \frac{dx_1 \, dx_2 \, \ldots \, dx_n \ N(x_1,...,x_n)}{D(x_1,...,x_n)}\,,
\end{align} 
where $N$ and $D$ are polynomials (or certain algebraic functions) in the $x_i$, that only has logarithmic singularities in the sense of (\ref{eq:dlog_general}), one should in principle be able to find an appropriate change of variables $x_i\mapsto g_i (x_j)$ to bring $\Omega$ into a manifest $\dlog$ representation (finding the primitive)
\begin{align}
\label{eq:dlog_sum}
\Omega = \sum_{k} c_k\ \dlog g^{(k)}_1 \, \cdots \, \dlog g^{(k)}_n\,, \qquad \text{where }  \dlog x \equiv \frac{dx}{x}\,.
\end{align}
The coefficients $c_k$ are the \emph{leading singularities} of the $n-$form $\Omega$ obtained by taking the maximal codimension-$n$ residue around $g^{(k)}_i=0$. Besides the logarithmic singularity property at the integrand level a la Eq.\eqref{eq:dlog_general} (associated with a preferred choice of master integrals that leads to canonical differential equations), the knowledge of an explicit $\dlog$ representation of the integrand (as in Eq.\eqref{eq:dlog_sum}) has not played a crucial role for the integration process. In the examples that we study here, however, knowing the manifest $\dlog$ form is the crucial initial step of our algorithm. 

In particular, we present a novel $\dlog$ differential equation which we apply to a number of Feynman integrals where we constructed the change of variables to write the $\dlog$-forms explicitly. As we will see, the $\dlog$ forms themselves play an important role for writing simple differential equations and reading off the symbol \cite{Goncharov:2010jf,Duhr:2011zq,Duhr:2012fh} of the Feynman integral. For one-loop and any planar integrals, we find it most convenient to work in the embedding space formalism \cite{Weinberg:2010fx} which is summarized in Appendix~\ref{app:embedding}. It is then straightforward to translate these formulae to any other appropriate kinematic space (loop momentum space, dual momentum space, momentum twistor space \cite{Hodges:2009hk}, etc.). The main advantage of our differential equations lies in the fact that the key steps (localization and generalized unitarity) are essentially independent of the number of scales involved in the problem. 

The remainder of this work is structured as follows; Section~\ref{sec:toymodels} illustrates the main ideas behind our differential equations on simple one and two-dimensional toy integrals. Section~\ref{sec:oneloop} applies these ideas to one-loop Feynman integrals where we rediscover the motivic formulae of Refs.~ \cite{Goncharov:1996tate,Spradlin:2011wp,Arkani-Hamed:2017ahv,Abreu:2017enx} from the point of view of $\dlog$ integrands. In Section~\ref{sec:rt}, we make use of residue theorems derived from an integrand perspective to reduce the number of independent directions of the differential equations. Effectively, this allows us to reduce the number of final entries of the symbol of the Feynman integral to a minimal set. In Section~\ref{sec:twoloop}, we extend the differential equations beyond one loop. We first discuss a simple one-dimensional toy example of a two-loop integral in subsection~\ref{subsec:twolooptoy} before presenting the $\dlog$ differential equation for the two-loop off-shell ladder integral in subsection~\ref{subsec:twoloopladder}. We end with our conclusions and an outlook to future work in Section~\ref{sec:conclusions}. A review of the embedding space formalism and and a discussion of $\dlog$ integrals on and off the null cone are included in Appendix~\ref{app:embedding}.

%
\section{Toy examples for $\dlog$ localization}
\label{sec:toymodels}
%

\subsection{Single-variable example}
\label{subsec:onevtoymodels}
To illustrate some of the ideas in a simple setting, consider the following integral
 \begin{equation}
   I(a,b) =  \int\limits^a_0  \frac{dx}{x+b} = \log \left(\frac{a+b}{b}\right)\,,
 \end{equation}
which everyone knows how to evaluate by finding the primitive of the integrand, $\dlog(x+b)$. An almost equivalent way of finding this result is to consider the differential of $I(a,b)$ as given by the Leibniz rule,
\begin{equation}
\label{eq:SingleVarEx}
dI(a,b) = \frac{da}{a+b}  +  \left[\frac{db}{a+b}-\frac{db}{b}\right] =\dlog \left(\frac{a+b}{b}\right)\,,
\end{equation}
where the $da$ term arises from the variation of the integration cycle and the $db$ term from the differential of the integrand, $\dlog(x+b)$. Alternatively, we could have changed variables to $y = x/a$, so that the full differential comes from the integrand. From this perspective, it is convenient to write the integral as follows
\begin{equation}
  I(a,b) = \int \limits^1_0 \dlog (y a+b) \,.
 \end{equation}
The fact that $\dlog (y a+b)$ is a closed form on the full space of $(y,a,b)$, implies the following relation
\begin{align}
  d^2\log (y a +b) =0 \quad \Rightarrow \quad d_{a,b} \big(d_y\log (y a +b)\big) = - d_y \big(d_{a,b}\log (y a +b)\big)
\end{align}
which is nothing but the statement that $d^2=0$ or that partial derivatives commute. Using this relation, we see that the differential of $I(a,b)$ arises purely from a boundary term by the fundamental theorem of calculus (``differentiation is the inverse of integration''). This seemingly pointless exercise provides us with some intuition about how to attack more complicated integrals whose integrands have a $\dlog$ form. 

\subsection{Multi-variable example}

To continue building up our intuition, we consider the following two-dimensional integral which depends on a parameter $a\in \mathbb{R}^+$
\begin{align}
  I(a) = \frac{1}{2\pi i}\int\limits_{\Sigma} \omega = \frac{1}{2\pi i}\int\limits_{\Sigma} dz d\zb \frac{\sqrt{1+4a}}{ (z \zb + a)[(z+1)(\zb +1) + a]}\,,
  \label{eq:toyrat}
\end{align}
where the volume form, $dz\,d\zb$, is oriented (i.e. $dz\wedge d\zb$) but we leave wedge products implicit throughout. The integration cycle, $\Sigma$, is taken to be (the compactification of) the real cycle in $\mathbb{C}^2$ where $z$ and $\zb $ are complex conjugate. One can easily evaluate this integral directly by using  polar coordinates $z = r\, e^{i \phi}$ and $\zb = r\, e^{-i \phi}$, with the following result
\begin{equation}
  I(a) = \log \left( \frac{1+2 a+\sqrt{1+4 a}}{1+2 a-\sqrt{1+4 a}} \right) =  \log \left( \frac{\sqrt{1+4 a}+1}{\sqrt{1+4 a}-1} \right)^2 \,. 
  \label{eq:toyres}
\end{equation}
We would like to reproduce the result in Eq.\eqref{eq:toyres} via a different strategy that generalizes to more complicated integrals and sheds light on some interesting features.

First, we rewrite the integrand, $\omega$, in \eqref{eq:toyrat} in $\dlog$ form 
\begin{equation}
  \omega = \frac12 \, \dlog \frac{z \zb + a}{(z+1)(\zb +1) + a} \, \dlog \frac{(z-z_+) (\zb-\zb_+)}{(z-z_-) (\zb-\zb_-)}\,,
  \label{eq:toydlog}
\end{equation}
where $(z_+,\zb_+)$ and $(z_-,\zb_-)$ are the two solutions to the following equations
\begin{align}
\label{eq:toycut}
  z \zb + a =0\,,
  \quad  (z+1)(\zb +1) + a =0\,,
\end{align}
More explicitly, we find
\begin{align}
\label{eq:zpm}
z_\pm = \zb_\mp=-\frac{1}{2} \left(1\pm\sqrt{1 + 4 a}\right)\,.
\end{align}
It is easy to check the equivalence between the $\dlog$ form in Eq.\eqref{eq:toydlog} and the rational form in Eq.\eqref{eq:toyrat} by using the chain rule, and the usual rules for wedge products.

\begin{figure}[t]
  \centering
  \raisebox{7pt}{\includegraphics[width=0.5\textwidth]{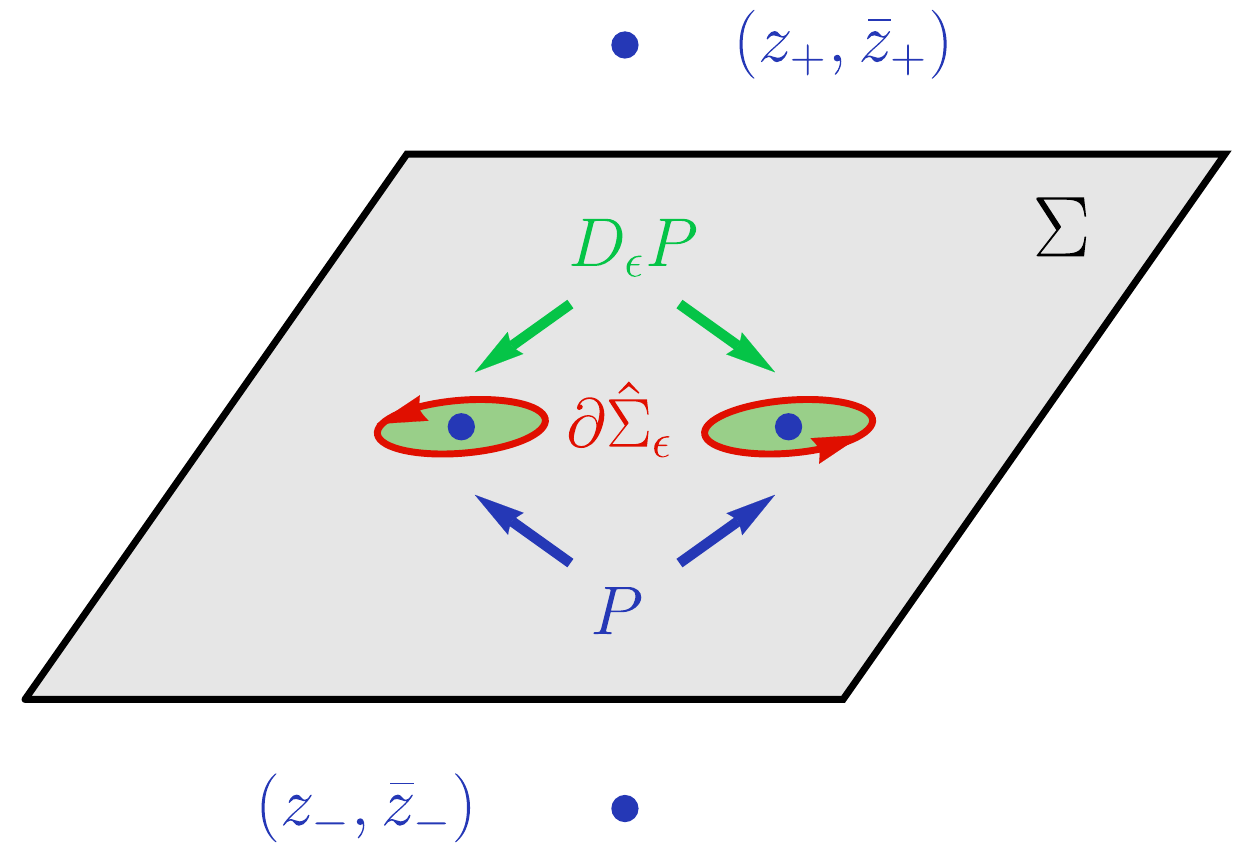}}
\hspace{0.5cm}
  \includegraphics[width=0.4\textwidth]{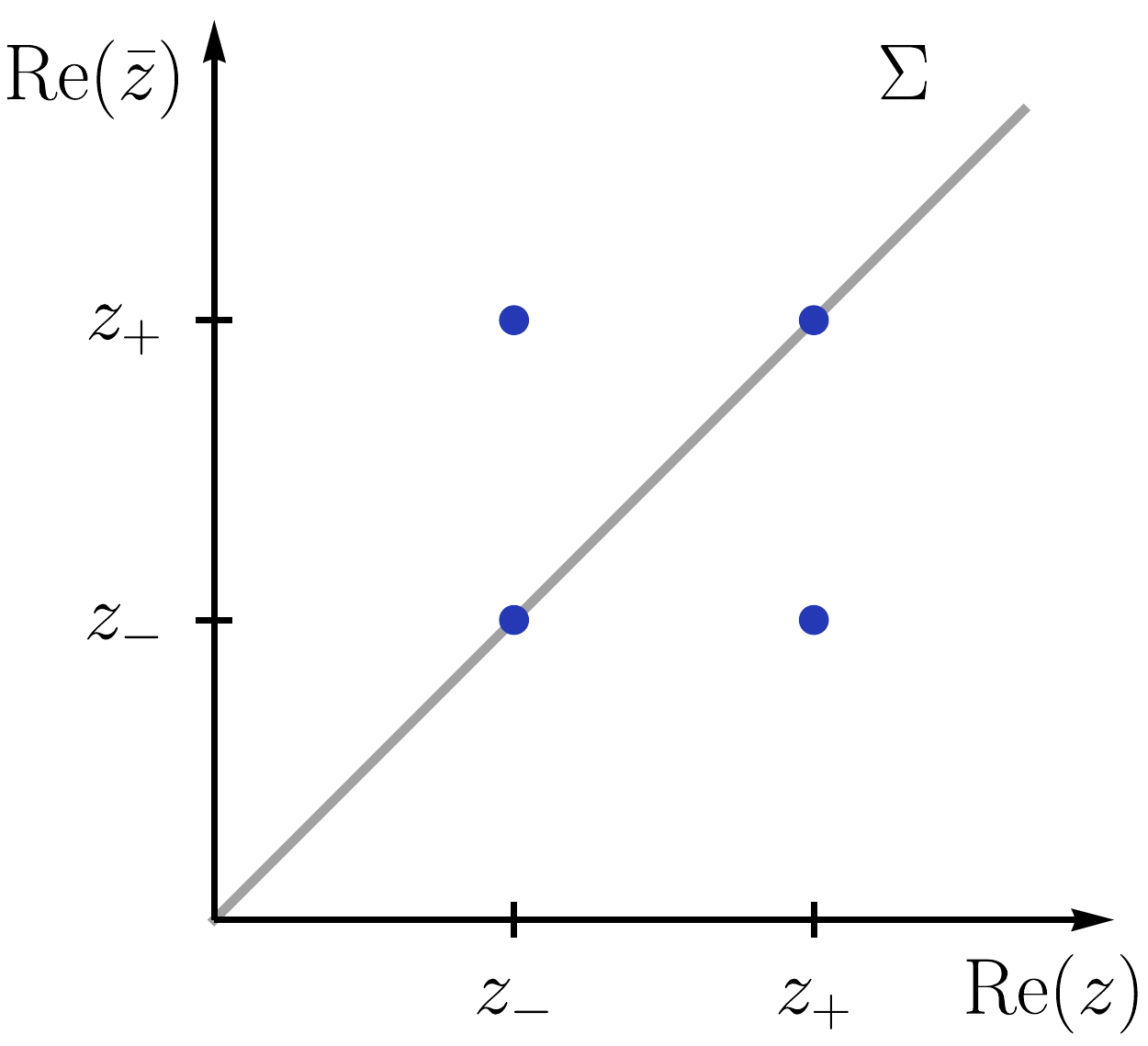}
  \caption{Sketch of the geometry of the various points and surfaces involved in the localization of the toy integral. In the left figure $\Sigma$ denotes the integration cycle, the green disks $D_\epsilon P$ denote the excision of the singularities $P$. The two points $(z_{\pm},\zb_{\pm})$ outside of the integration cycle $\Sigma$ are the solutions to Eq.\eqref{eq:toycut}. The right figure shows the relative positions of all the points and $\Sigma$ in the subspace ${\rm Im}(z) = {\rm Im}(\zb) = 0$.}
  \label{fig:bubblegeo}
\end{figure}

The $\dlog$ form in Eq.\eqref{eq:toydlog} suggests that $\omega$ could be considered in the full space of $z,\zb$ and also $a$, so that the total differential reads
\begin{equation}
  d = d_i + d_a \,,
\end{equation}
where $d_i$ ($i$ for integration) denotes the differential in the direction of the integration variables $z,\zb$.
From this perspective we can decompose the two-form $\omega$
\begin{equation}
  \omega = \omega^{(2,0)} + \omega^{(1,1)}\,,
\end{equation}
into components labeled by superscripts $(r_i,r_e)$ indicating the number of differentials in the integration ($r_i$) and external variables ($r_e$). In this language, the integrand in Eq.\eqref{eq:toyrat} would be labeled as $\omega^{(2,0)}$, and
\begin{equation}
\omega^{(1,1)} = \frac{1+2 a (2+z+3 \zb)+2  \zb-(z-\zb) (2 z \zb+z+\zb)}{2 \sqrt{1+4 a}\,  [z -z_+][z -z_-] [z \zb-a] [(z+1)(\zb +1) + a]} dz\,da +  (z\leftrightarrow \zb)\,.
\label{eq:toy11form}
\end{equation}
Note that in the calculation of an integral the integration cycle picks out the correct component of $\omega$. Viewing $\omega$ as a form on the combined space of integration variables and external parameters has the advantage that $\omega$ is a closed form on the full space, i.e. $d\omega=0$. This is easy to see from Eq.\eqref{eq:toydlog} and the familiar identity $d^2=0$. In terms of $\omega$'s components, $d\omega=0$ implies the following relation
\begin{equation}
  d\omega = 0 \quad \leftrightarrow \quad d_a \omega^{(2,0)} = - d_i \omega^{(1,1)}\,.
  \label{eq:drel}
\end{equation}
The trivial looking relation in Eq.\eqref{eq:drel} has far-reaching consequences when trying to derive a differential equation for $I(a)$ in $a$. Looking at the rational form in Eq.\eqref{eq:toyrat} one would naively conclude that the result of taking derivatives in $a$ is a similar integral with double poles. On the other hand, Eq.\eqref{eq:drel} reveals that the resulting form will be a total derivative in the integration variables $z,\zb$. That being said, how is it possible that $I(a)$ is non-zero if it is the integral of a total derivative? At this point it is crucial to realize that both the $\dlog$ form Eq.\eqref{eq:toydlog} as well as the form in Eq.\eqref{eq:toy11form} contain additional poles, $(z-z_\pm)$ and  $(\zb-\zb_\pm)$, that are naively not present in the original form of the integrand Eq.\eqref{eq:toyrat}. Whereas $\omega^{(2,0)}$ does not have a singularity at these poles, i.e. the residue at these poles is zero, the residue of $\omega^{(1,1)}$ is non-vanishing. Furthermore, the singularities at 
\begin{equation}
  \label{eq:Ppoints}
  P\equiv \{(z,\zb) = (z_+,\zb_-) \, \cup \, (z,\zb) = (z_-,\zb_+)\}
\end{equation}
intersect the integration cycle $\Sigma$. Thus, in order to make use of Eq.\eqref{eq:drel}, we must excise these singularities as illustrated in Fig.~\ref{fig:bubblegeo}.
More explicitly, the resulting cycle is
\begin{equation}
  \hat\Sigma_\epsilon = \Sigma \, / \,D_{\epsilon}P \,.
\end{equation}
where $D_{\epsilon}P$ is a small disk of radius $\epsilon$ containing the singularities at P\footnote{For a precise definitions of higher dimensional residues in the context of Feynman integrals, see~\cite{Abreu:2017ptx}}.
At the end of the day, the total derivative localizes to a boundary term by Stokes theorem
\begin{equation}
  d I = - \frac{1}{2\pi i} \int\limits_{\hat\Sigma_\epsilon} d_i\omega^{(1,1)} = - \frac{1}{2\pi i} \int\limits_{\partial \hat\Sigma_\epsilon} \omega^{(1,1)}\,.
\end{equation}
Obviously, this reduces the dimension of the integration by one. But notice that $\partial \hat\Sigma_\epsilon$ is comprised of little circles surrounding the singular points $P$, so that the left-over integral is given by residues of $\omega^{(1,1)}$ 
\begin{equation}
  d I = - {\rm Res}_P[\omega^{(1,1)}] \,.
\end{equation}
These can be readily evaluated from the $\dlog$ form \eqref{eq:toydlog} or \eqref{eq:toy11form}, yielding
\begin{equation}
  d I =  \dlog  \frac{[z_+ \zb_- + a][(z_-+1)(\zb_+ +1) + a]}{[z_- \zb_+ + a][(z_++1)(\zb_- +1) + a]} \,.
  \label{eq:locres}
\end{equation}
Plugging in the values of $z_{\pm}$ and $\zb_{\pm}$ in Eq.\eqref{eq:zpm} one can check that this reproduces the result in Eq.\eqref{eq:toyres}.

After this song and dance with simple toy examples, the reader might wonder what any of this has to do with Feynman integrals? In fact, it turns out that the example we just worked through is secretly the massive bubble integral in $D=2$, whose representation in ordinary loop momentum variables is
\begin{equation}
	\raisebox{-27pt}{\includegraphics[scale=.5, trim= -20 0 0 0]{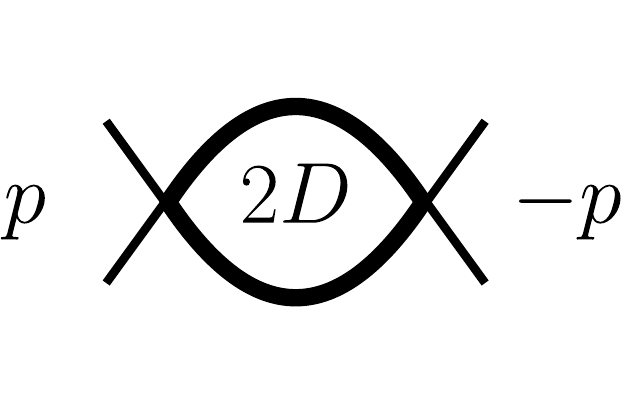}} \hspace{5pt} = \int \frac{d^2\ell}{i\pi} \frac{\sqrt{p^2 (p^2-4m^2)}}{[\ell^2-m^2][(\ell+p)^2-m^2]}\,.
\end{equation}
The form in Eq.\eqref{eq:toyrat} is obtained by choosing light-cone variables
\begin{equation}
  \ell^0-\ell^1 = z\, (p^0-p^1)\,, \quad \ell^0+\ell^1 = \zb\,(p^0+p^1)\,,
\end{equation}
and setting $a= m^2/(-p^2)$. The integration over the real cycle in Eq.\eqref{eq:toyrat} corresponds to integrating over Euclidean loop momenta, after Wick rotation. The $\dlog$ form of the $2D$ bubble follows a more familiar structure \cite{ArkaniHamed:2012nw}
\begin{equation}
    \omega=\frac12 \, \dlog \frac{\ell^2-m^2}{(\ell+p)^2-m^2} \, \dlog \frac{(\ell-\ell_+)^2}{(\ell-\ell_-)^2}\,,
\end{equation}
where $\ell_\pm$ are the two solutions to the maximal cut. The $+$ and $-$ subscripts denote the fact that the residues of the integrand at these points are $(+1,-1)$ respectively\footnote{This is also true for the higher dimensional generalizations of this integral discussed in later sections. For more details see e.g.~\cite{ArkaniHamed:2010gh}.}.  Finally, the result of the localization in Eq.\eqref{eq:locres} can be written compactly as
\begin{equation}
  d \left[\!\!\!\!\!\raisebox{-27pt}{\includegraphics[scale=.5, trim= -20 0 0 0]{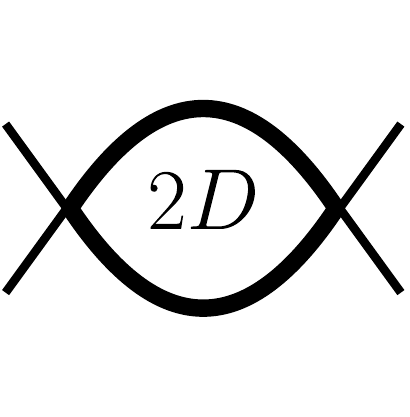}}\right] = \dlog \frac{[\ell_{\bullet}^2-m^2][(\ell_{\circ}+p)^2-m^2]}{[\ell_{\circ}^2-m^2][(\ell_{\bullet}+p)^2-m^2]} \,.
\end{equation}
Where the $\ell_{\circ},\ell_{\bullet}$ are the two solutions to $(\ell-\ell_+)^2=(\ell-\ell_-)^2=0$. In the following sections, we will see that this localization procedure can be applied to more complicated Feynman integrals at one and two loops.

%
\section{One loop examples}
\label{sec:oneloop}
%

In this section we work through a simple one-loop example, illustrating the localization of $\dlog$ integrals that give rise to a recursive differential structure. Towards the end, in subsection~\ref{subsec:dgonddim}, we give a general formula for the differential of scalar $D$-gon integrals in $D$ dimensions which is reminiscent of \emph{motivic formulae} that have been derived \cite{Goncharov:1996tate,Spradlin:2011wp,Arkani-Hamed:2017ahv,Abreu:2017enx} in the literature.  Amusingly, these differential equations also appear in the context of volumes of hyperbolic simplices \cite{aomoto1977,Aomoto:1992,Davydychev:1997wa,Mason:2010pg,Nandan:2013ip}  and were known to Schl\"afli in the 19$^{\text{th}}$ century \cite{Schlaefli:1860}.

\subsection{Box integral with internal masses}

As a first example, we study the one-loop box integral with internal masses and all massless external legs. 
\begin{equation}
  I_4(s,t,m^2) = \includegraphics[scale=.5, trim= 10 70 0 0]{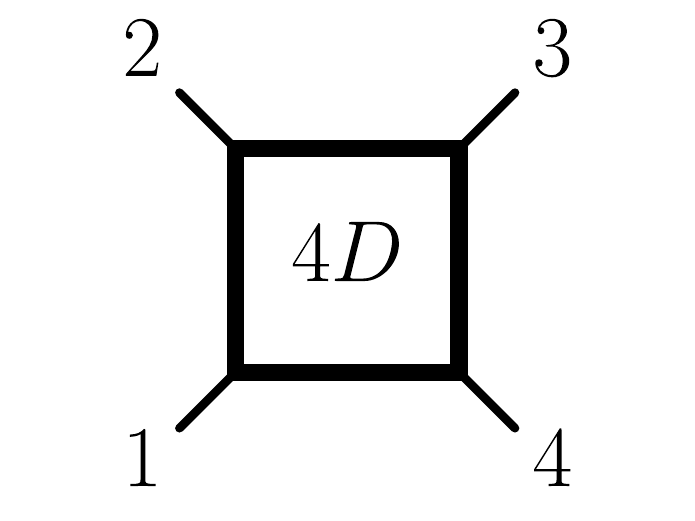}
  \vspace{25pt}
\end{equation}
This integral has been described in \cite{Davydychev:1993ut} and plays a prominent role in the study of Coulomb branch amplitudes in $\N=4$ SYM theory \cite{Caron-Huot:2014lda,Henn:2014qga}.
In momentum space the integral is given by,
\begin{align}
\label{eq:0massBox_internalMasses}
I_4 = \int \frac{d^4 \ell}{i\pi^2}\frac{\sqrt{st(st - 4m^2 (s+t))}}{[\ell^2-m^2] [(\ell+p_1)^2-m^2][(\ell+p_1+p_2)^2-m^2][(\ell-p_4)^2-m^2]}\,,
 \end{align}
where $s=(p_1+p_2)^2$ and $t=(p_1+p_4)^2$ are the usual Mandelstam variables.

From here on, we work in the embedding space formalism \cite{Dirac:1936fq,Weinberg:2010fx}, reviewed in Appendix~\ref{app:embedding}. Let us emphasize that none of the following steps rely on the existence of an embedding space and can likewise be performed in loop-momentum space, so that our procedure applies to more general integrals (including non-planar). We only pass to embedding space for technical simplicity and notational clarity. 
 
In terms of the momenta above, the coordinates of the external kinematics in embedding space are given by \cite{Caron-Huot:2014lda}
 \begin{equation}
   X_1 = \begin{pmatrix} 0^\mu \\ m^2 \\ 1 \end{pmatrix} \,, \quad 
   X_2 = \begin{pmatrix} -p_1^\mu \\ m^2 \\ 1 \end{pmatrix} \,, \quad
   X_3 = \begin{pmatrix} -(p_1+p_2)^\mu \\ -s+m^2 \\ 1 \end{pmatrix} \,, \quad
   X_4 = \begin{pmatrix} p_4^\mu \\ m^2 \\ 1 \end{pmatrix} \,, 
 \end{equation}
and the loop momentum corresponds to
 \begin{equation}
   Y = \begin{pmatrix} \ell^\mu \\ -\ell^2 \\ 1 \end{pmatrix} \,.
 \end{equation}
In these variables, taking into account the rules summarized in Appendix~\ref{app:embedding}, the integral $I_4$ takes the simple form
\begin{align}
I_4 = \int\limits_{\Sigma_4} \frac{ \langle Y d^5Y\rangle \sqrt{-\det(X_iX_j) }}{(YX_1)(YX_2)(YX_3)(YX_4)}\,,
  \label{eq:massiveboxrat}
\end{align}
where $\det(X_iX_j)$ denotes the Gram-determinant of the external points, and $\Sigma_4$ denotes the integration cycle $(YY)=0$. The $\dlog$ form in embedding space is extremely simple
\begin{align}
  I_4 = \int\limits_{\Sigma_4}  \frac12 \dlog\left(\frac{YX_1}{Y X_2}\right)\dlog\left(\frac{YX_2}{Y X_3}\right)\dlog\left(\frac{YX_3}{Y X_4}\right)\dlog\left(\frac{YX_{+}}{Y X_-}\right)\,,
  \label{eq:massiveboxdlog}
\end{align}
and $X_\pm$ denote the two solutions to the maximal cut of the box. As will be shown later, often, we do not even need the explicit form of $X_\pm$. If explicit solutions are required, a convenient way of finding $X_\pm$ proceeds by choosing a parameterization of $Y$ in terms of six points, the four points $X_{i=1,\ldots,4}$ in the box integral and two additional generic points $X_5,X_6$
\begin{equation}
  Y = \sum^6_{i=1} a_i X_i\,.
  \label{eq:embeddingLoop}
\end{equation}
In these variables, the cut conditions and the condition\footnote{$I$ denotes the point at infinity, which is described in Appendix~\ref{app:embedding}.} $(YI)=1$ are a set of linear equations which are straightforward to solve. Finally, one has to impose the quadratic equation $(YY)=0$ to land on the integration contour $\Sigma_4$.  In terms of the $a_i$ variables it is easy to check the equivalence of Eqs.~\eqref{eq:massiveboxdlog} and \eqref{eq:massiveboxrat}.

\subsubsection{Localization and generalized unitarity}
%
\begin{figure}[t]
  \centering
  \includegraphics[width=0.52\textwidth]{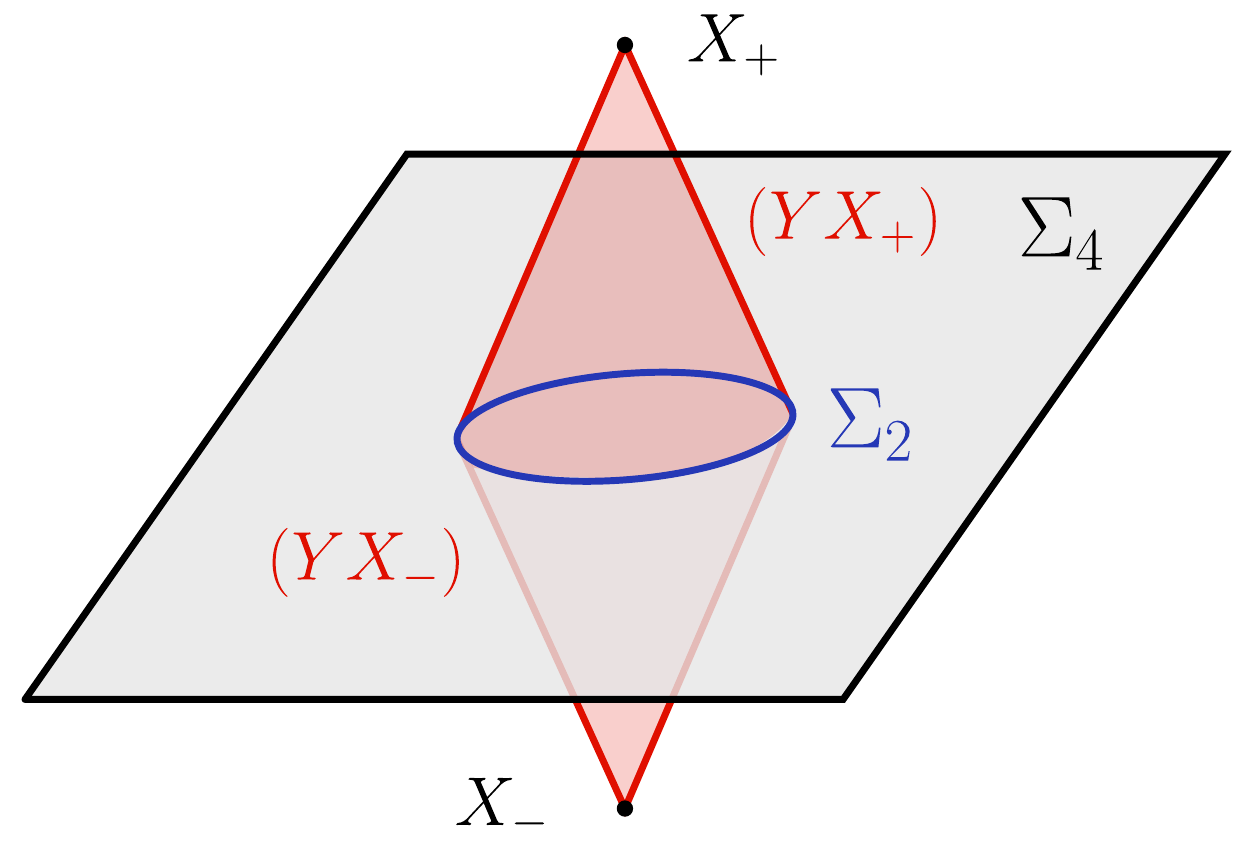} \hspace{0.5cm}
  \caption{Illustration of the geometry of the intersection of the integration cycle $\Sigma_4$ and the light-cones $(YX_{\pm})=0$. $\Sigma_2$ is the cycle that the differential of the $\dlog$ integral localizes to.}
  \label{fig:boxgeometry}
\end{figure}
We would like to proceed along the lines of the $D=2$ bubble-example of the previous section. Recall the key step: the differential of the integral localizes to a codimension two cycle through the application of Stokes theorem and a residue integral. In the previous case this cycle was just the set of points $P$ in Eq.\eqref{eq:Ppoints}. More generally, the localization surface is the intersection of the real integration cycle (i.e. Euclidean loop momenta) and the new singularities in the $\dlog$ form, namely $(YX_+)=0$ and $(YX_-)=0$. In the case at hand such singularities appear in the $(3,1)$ component of the $\dlog$ form \eqref{eq:massiveboxdlog}. The geometry of this setup is illustrated in Fig.~\ref{fig:boxgeometry}. We call the localization surface $\Sigma_2$. Working in embedding space makes it clear that $\Sigma_2$, being the intersection of the original quadric $\Sigma_4$ with the two hyperplanes $(YX_\pm)=0$, is just another quadric. Any two quadrics in projective space are related by a conformal transformation. This implies that we can interpret the final integration cycle $\Sigma_2$ as just the ordinary Feynman contour in $D{=}2$. The result of the localization procedure in the case of the massive box integral is
\begin{align}
  \label{eq:massiveboxloc}
  dI_4 = \int\limits_{\Sigma_2} \Omega^{(2,1)}= \int\limits_{\Sigma_2}   \dlog\left(\frac{YX_1}{Y X_2}\right)\dlog\left(\frac{YX_2}{Y X_3}\right)\dlog\left(\frac{YX_3}{Y X_4}\right)\,,
\end{align}
where it is clear that one of the differentials needs to be in the direction of external variables for the expression to make sense. 

Our goal is to derive a differential equation of the schematic form 
\begin{equation}
  dI_4 = \sum_i \dlog\alpha_i \, I^{(i)} 
  \vspace{-.3cm}
\end{equation}
where $\alpha_i$ are functions of external variables only and $I^{(i)}$ are integrals over $\Sigma_2$. Eq.\eqref{eq:massiveboxloc} is not manifestly of this form and requires some extra manipulations. A key feature of the integrand, $\Omega^{(2,1)}$ in Eq.\eqref{eq:massiveboxloc} is that it only has the original propagator-type singularities that were already present in the rational form Eq.\eqref{eq:massiveboxrat}. With the help of generalized unitarity \cite{Bern:1994cg,Bern:1997sc,Britto:2004nc,Bern:2007ct}, we can then bring $\Omega^{(2,1)}$ to the desired form. In our context, by generalized unitarity we simply mean the reconstruction of a rational form by matching its poles and residues. This is usually done by starting with an ansatz in terms of basis of local integrals with undetermined coefficients, which are then fixed by calculating residues. Equivalently, we will refer to this step as \emph{partial fractioning} the form $\Omega^{(2,1)}$. From the perspective of unitarity, a sufficient and complete basis of $D=2$ integrands with only logarithmic singularities and no pole at infinity (due to the absence of poles in $(YI)$) is given by a set of parity-odd triangles and all scalar bubbles
\vspace{-.8cm}
\begin{align}
  I_2^s &= \raisebox{-35pt}{\includegraphics[scale=.5]{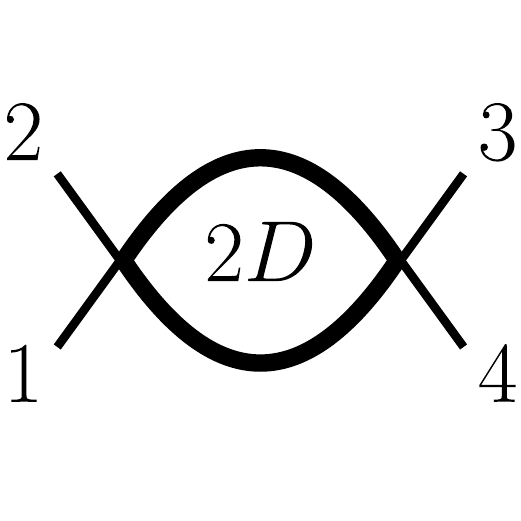}}\,, \quad 
  I_2^1 = \raisebox{-42pt}{\includegraphics[scale=.5]{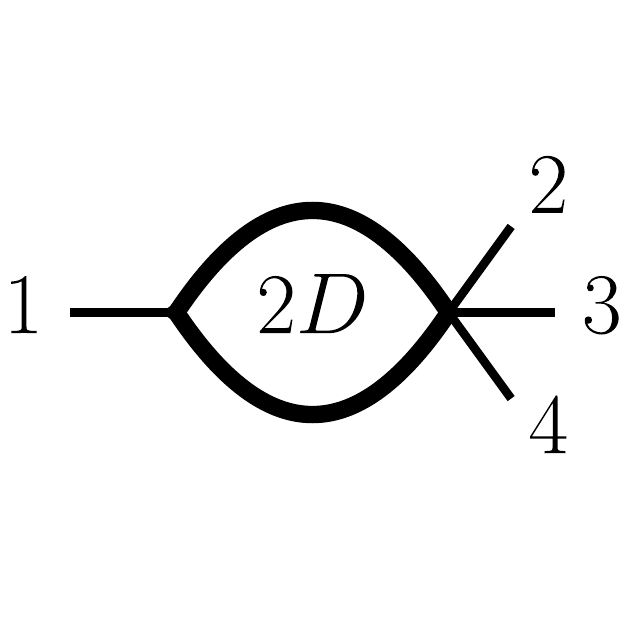}}\,, \quad 
  I_2^2 = \raisebox{-42pt}{\includegraphics[scale=.5]{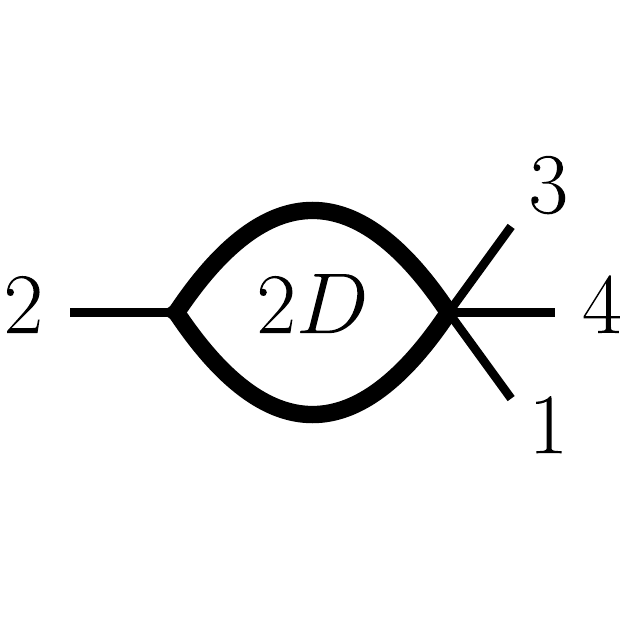}}\,, \\[-34pt]
  I_2^t &= \raisebox{-35pt}{\includegraphics[scale=.5]{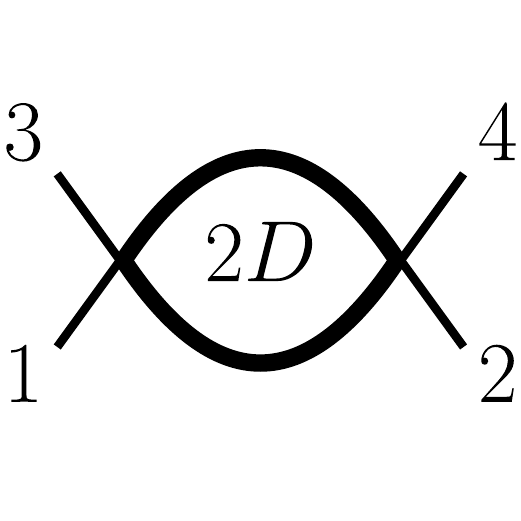}}\,, \quad 
  I_2^3 = \raisebox{-42pt}{\includegraphics[scale=.5]{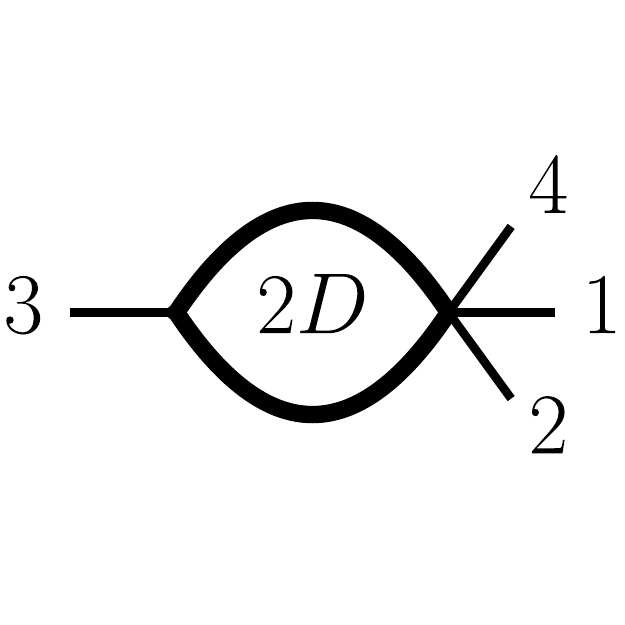}}\,, \quad 
  I_2^4 = \raisebox{-42pt}{\includegraphics[scale=.5]{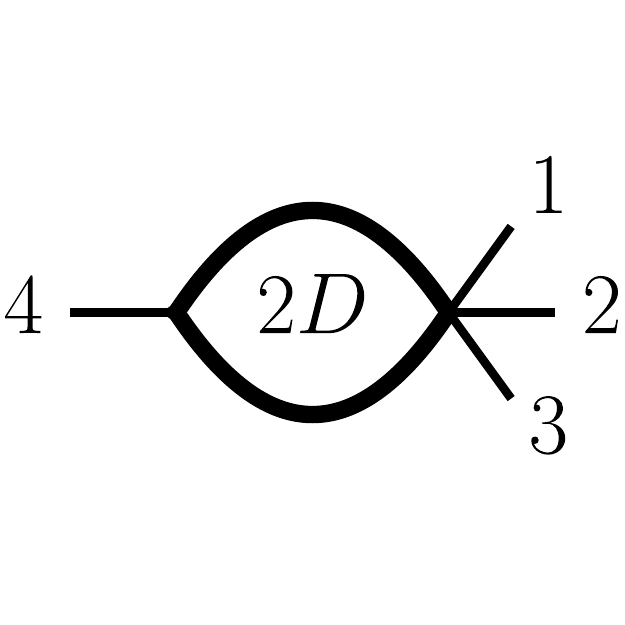}}\,. 
\end{align}
\vskip -.8cm\noindent
We will not concern ourselves with the former since they integrate to zero\footnote{Unlike at one loop, parity-odd contributions in a single loop of a higher-loop integrand cannot be neglected. These play an important role for our two-loop example in Sec.~\ref{sec:twoloop}.}.  
Thus the differential equation takes the form
\begin{align}
  \label{eq:dI4basis}
  dI_4 &= c_s I_2^s +  c_t I_2^t + 
  c_{1} I_2^1 +  c_{2} I_2^2 +  c_{3} I_2^3 + 
  c_{4} I_2^4\,.
\end{align}
The coefficient of any scalar bubble can be determined by a residue computation. As an example, let us calculate explicitly the coefficient of the bubble with propagators $(YX_1)(YX_2)$. The residue can be extracted from the $\dlog$ form in Eq.\eqref{eq:massiveboxdlog} by picking out the appropriate piece
\begin{equation}
  \Omega^{(2,1)} = \dlog(YX_1)\dlog(YX_2)\dlog\left(\frac{YX_3}{Y X_4}\right) +\cdots\,.
\end{equation}
It is clear that the coefficient of the bubble in Eq.\eqref{eq:dI4basis} is
\begin{equation}
  c_{1} = \frac12 \left({\rm Res}[\Omega^{(2,1)},X_+^{34}] -  {\rm Res}[\Omega^{(2,1)}, X_-^{34}]\right) = \frac12 \dlog \frac{(X_+^{34}X_3)(X_-^{34} X_4)}{(X_+^{34} X_4)(X_-^{34}X_3)}\,,
  \label{eq:excoeff}
\end{equation}
where $X_\pm^{34}$ are the solutions to cutting $(YX_1)$ and $(YX_2)$ on the support of $\Sigma_2$\footnote{The superscripts in the notation indicate the \emph{uncut} propagators. This will make the generalization to higher-dimensional integrals more concise.}. From the $D=4$ perspective these points are the solution to the cut equations
\begin{equation}
  (YX_1)=(YX_2)=(YX_+)=(YX_-) = 0\,.
\end{equation}
The factor of $1/2$ in Eq.\eqref{eq:excoeff} is familiar from unitarity and due to extracting parity even combinations; the minus sign might be more unfamiliar but stems from calculating the residues carefully (that is, including the appropriate Jacobian). For calculating the residues of the $\dlog$ forms, it is often useful to manipulate the arguments of the $\dlog$s using the following identities,
\begin{align}
\begin{split}
\cdots \dlog a \cdots \dlog b \cdots & = \cdots \dlog a \cdots \dlog \frac{b}{a} \cdots   
						      = \cdots \dlog a \cdots \dlog (a\, b) \cdots\,,
\end{split}						
\label{eq:dlogidentities}
\end{align}
which follow trivially from the antisymmetry of the wedge product.

In order to evaluate the argument of the $\dlog$ in Eq.\eqref{eq:excoeff} in terms of explicit external variables, such as masses and Mandelstam invariants, one can evaluate the solution to the cut using one's favorite parameterization of the loop variables. Alternatively, in embedding space there is a trick to evaluate them without solving the cuts explicitly by using the completeness relations described in Appendix~\ref{app:embedding}. We find that the only non-vanishing coefficients are $c_{s}$ and $c_{t}$ and the differential equation reads
\begin{align}
  dI_4 &= \frac12 \dlog \frac{(X_+^{24}X_2)(X_-^{24} X_4)}{(X_+^{24} X_4)(X_-^{24}X_2)} \, I_2^s + \frac12 \dlog \frac{(X_+^{13}X_1)(X_-^{13} X_3)}{(X_+^{13} X_3)(X_-^{13}X_1)} \, I_2^t  \\
&= \frac12 \dlog \left(\frac{\beta_{uv}-\beta_u}{\beta_{uv}+\beta_u}\right)^2 \, I_2^s + \frac12 \dlog \left(\frac{\beta_{uv}-\beta_v}{\beta_{uv}+\beta_v}\right)^2 \, I_2^t
\label{eq:diffbox}
\end{align}
where $\beta_u = \sqrt{1+\frac{4m^2}{-s}}$, $\beta_v = \sqrt{1+\frac{4m^2}{-t}}$ and $\beta_{uv} = \sqrt{1+\frac{4m^2}{-s} + \frac{4m^2}{-t}}$.

The result for the massive $D=2$ bubbles, $I_2^s$ and $I_2^t$, can be obtained from the discussion in the previous section by choosing $a=\frac{m^2}{-s}$ or $a=\frac{m^2}{-t}$ in Eq.\eqref{eq:toyres} respectively. Since the differential equations relate integrals in dimensions differing by two, we can interpret the differential equations for the bubbles as relating them to the trivial integral in $D=0$, $I_0$, which is just a constant that we normalize to one. Combining our results we find that the set of integrals satisfies the following system of differential equations
\begin{equation}
  \label{eq:ourdiff}
  d\begin{pmatrix} I_4  \\[10pt] I_2^s \\[10pt] I_2^t \\[10pt] I_0 \end{pmatrix} 
  =   \begin{pmatrix} 
  	0 & \frac12 \dlog\left(\frac{\beta_{uv}-\beta_u}{\beta_{uv}+\beta_u}\right)^2 & \frac12 \dlog\left(\frac{\beta_{uv}-\beta_v}{\beta_{uv}+\beta_v}\right)^2 & 0 \\
   	0 & 0 & 0 & \dlog\left(\frac{\beta_{u}+1}{\beta_{u}-1}\right)^2 \\  
	0 & 0 & 0 & \dlog\left(\frac{\beta_{v}+1}{\beta_{v}-1}\right)^2 \\[10pt] 
	0 & 0 & 0 & 0 
      \end{pmatrix} 
\begin{pmatrix} I_4  \\[10pt] I_2^s \\[10pt] I_2^t \\[10pt] I_0 \end{pmatrix}
\end{equation}
%

\subsubsection{Relation to canonical differential equations}

In \cite{Caron-Huot:2014lda}, Caron-Huot and Henn obtained a similar result using more standard differential equation methods. The chosen basis of master integrals includes 
\begin{align}
  \begin{split}
    g_1 &= 2m^2 G^4_{0,0,0,3} = 2m^2 G^4_{0,0,3,0} = 2m^2 G^4_{0,3,0,0} = 2m^2 G^4_{3,0,0,0} \,,\\
    g_2 &= -\sqrt{s(s-4m^2)} G^4_{1,0,2,0}\,, \\
    g_3 &= -\sqrt{s(s-4m^2)} G^4_{0,1,0,2}\,, \\
    g_6 &= \frac{1}{2} \sqrt{st(st-4m^2(s+t))} G^4_{1,1,1,1}\,,
   \end{split}
  \label{eq:johannesmasters}
\end{align}
where
\begin{equation}
  G^4_{a,b,c,d} = \int\frac{d^4\ell}{i \pi^2} \frac{1}{[\ell^2]^a[(\ell-k_1)^2]^b [(\ell-k_{12})^2]^c [(\ell+k_4)^2)]^d }\,.
\end{equation}
They found the following differential equation in $D=4$ by taking the $\epsilon\rightarrow 0$ limit of the differential equation in canonical $\epsilon$ form \cite{Henn,Henn:2014qga}
\begin{equation}
  \label{eq:theirdiff}
  d\begin{pmatrix} g_6  \\[10pt] g_3 \\[10pt] g_2 \\[10pt] g_1 \end{pmatrix} 
  	=  \begin{pmatrix} 
		0 &  \dlog\left(\frac{\beta_{uv}-\beta_v}{\beta_{uv}+\beta_v}\right) & \frac12 \dlog\left(\frac{\beta_{uv}-\beta_u}{\beta_{uv}+\beta_u}\right) & 0 \\ 
		0 & 0 & 0 & \dlog\left(\frac{\beta_{v}-1}{\beta_{v}+1}\right) \\  
		0 & 0 & 0 & \dlog\left(\frac{\beta_{u}-1}{\beta_{u}+1}\right) \\[10pt] 
		0 & 0 & 0 & 0
	   \end{pmatrix} 
  \begin{pmatrix} g_6  \\[10pt] g_3 \\[10pt] g_2 \\[10pt] g_1 \end{pmatrix}\,.
\end{equation}
The main difference between the two implementations is that the bubble integrals in Eq.\eqref{eq:johannesmasters} are four-dimensional integrals with doubled propagators. As a consistency check, one can use  dimension shifting identities \cite{Bern:1992em,Bern:1993kr,Tarasov:1996br} and integration by parts relations (IBP) relations \cite{Chetyrkin:1981qh,Tkachov:1981wb} to relate both bases of master integrals to one another. In the usual basis of scalar integrals the dimension shifting formula for the bubble is
\begin{equation}
  G^{D-2}_{1,0,1,0} = \frac{2(D-3)}{ -s + 4m^2}  G^D_{1,0,1,0} 
  - \frac{D-2}{m^2(-s + 4m^2)} G^D_{1,0,0,0} \,,
\end{equation}
and the IBP reduction of the integral relevant to $g_2$ is
\begin{equation}
  G^D_{1,0,2,0} = \frac{D-3}{-s + 4 m^2}     G^D_{1,0,1,0}
              - \frac{D-2}{2 m^2 (-s+4 m^2)} G^D_{1,0,0,0} \,.
\end{equation}
Comparing the equations above, we find that $ G^4_{1,0,2,0} = \frac12 G^{2}_{1,0,1,0} $, i.e., some of the integrals in the basis \eqref{eq:johannesmasters} are simply dimension-shifted bubbles. We conclude that both differential equations, \eqref{eq:ourdiff} and \eqref{eq:theirdiff}, are equivalent with the following identifications
\begin{align}
  \begin{split}
    g_6 &=   \frac12 \sqrt{st(st-4m^2(s+t))} G^{4}_{1,1,1,1} = \frac12 I_4 \,, \\
    g_3 &= -\frac12 \sqrt{s(s-4m^2)} G^{2}_{0,1,0,1} = -\frac{1}{2} I_2^t \,, \\
    g_2 &= -\frac12 \sqrt{s(s-4m^2)} G^{2}_{1,0,1,0} = -\frac{1}{2} I_2^s \,, \\
    g_1 &=  I_0 =1 \,.
   \end{split}
\end{align}
Our analysis suggests that a natural way of interpreting the canonical basis is as integrals shifted to the dimension where they are $\dlog$. We will see later that differential equations also produce mixed-dimension integrals at two loops.

\subsection{$D$-gons in $D$ dimensions}
\label{subsec:dgonddim}

The discussions of the previous two subsections immediately make clear that a similar procedure generalizes to more complicated one-loop integrals. It is apparent that general $D$-gon integrals in $D$ spacetime dimensions have the following $\dlog$ form
\begin{equation}
\label{eq:ngon_n_dimensions_dlog}
I^D = \frac12 \int\limits_{\Sigma_D} \dlog \frac{(YX_1)}{(YX_2)}\dlog\frac{(YX_2)}{(YX_3)} \cdots  \dlog \frac{(YX_{D-1})}{(YX_D)}\dlog  \frac{(YX_+)}{(YX_-)}\,,
\end{equation}
\vskip-.4cm\noindent
where $X_{\pm}$ are the solutions to the maximal cut equations $(YX_i) =0\,,  \forall i \in \{1,...,D\}$. Note that this $\dlog$ form is valid for arbitrary masses and external kinematic configurations.  For simplicity, we exclude special IR-divergent cases in our current discussion and leave their detailed study for future work. The localization and partial fraction procedures described in the previous two subsections immediately extend to the general $D$-gon case, for which we can now derive a differential equation
\begin{equation}
\label{eq:dgondiffeq}
dI^{(D)} = \frac{1}{2} \sum_{i<j} \dlog \frac{(X_iX^{ij}_+)}{(X_jX^{ij}_+)}\frac{(X_jX^{ij}_-)}{(X_iX^{ij}_-)}\, I^{(D-2)}_{ij} \equiv \frac{1}{2} \sum_{i< j} \dlog u_{ij}\, I^{(D-2)}_{ij}\,,
\end{equation} 
\vskip-.3cm\noindent
where $I^{(D-2)}_{ij}$ are scalar $(D-2)$-gon integrals in $D-2$ dimensions. The propagator structure of $I^{(D-2)}_{ij}$ is obtained from $I^{(D)}$ by pinching the two propagators $(YX_i)$ and $(YX_j)$.  The special points $X^{ij}_{\pm}$ appearing in the $\dlog$ of the cross-ration $u_{ij}$ in Eq.\eqref{eq:dgondiffeq} only depend on external kinematics (and masses). Explicitly, the points $X^{ij}_{\pm}$ are solutions to the localization and partial-fraction relations 
\begin{equation}
 X^{ij}_{\pm} \leftrightarrow (YX_+) = (YX_-) = 0\,, \quad (YX_a)=0\,, \quad\forall a \notin \{i,j\}\,.
\end{equation}
Comparing the structure of Eq.\eqref{eq:dgondiffeq} with the formulae first obtained by Goncharov \cite{Goncharov:1996tate} from mixed Tate motives and applied to Feynman integrals by Spradlin and Volovich \cite{Spradlin:2011wp}, we find a striking similarity. In \cite{Spradlin:2011wp}, the discussion focuses on $2m$-gon integrals in $2m$ dimensions in order to avoid complications from square roots in the denominator after Feynman parameter integrals. A similar formula has later been generalized in \cite{Arkani-Hamed:2017ahv} to general one-loop projective Feynman parameter integrals. The structure of the scalar Feynman parameter integrals is fully encoded in a particular quadric $Q$, that only depends on the external dual points (in the notation of \cite{Spradlin:2011wp}, these are the $x_i$)
\begin{align}
F_m(Q) = \Gamma(m) \int 
				\frac{\langle W d^{2m-1}W\rangle \sqrt{-\det Q}}{\left(\frac{1}{2} W\cdot Q\cdot W\right)^m}\,,
\end{align}
with the quadric defined by $Q_{ij} = (x_i-x_j)^2 \equiv x^2_{ij}$, and the standard holomorphic measure defined in Eq.\eqref{eq:holomorphic_measure}. From their analysis, they find a recursive structure of the function $F_m$ by acting with a total differential,
\begin{align}
  \label{eq:diffMotivic}
 dF_m(Q) = \half \sum\limits_{i<j} \dlog R_{ij} \ F_{m-1} (Q_{\bar i \bar j})\,,
\end{align}
where going from the quadric $Q_{ij}$ to the reduced quadric $Q_{\bar i \bar j}$ corresponds to removing rows and columns $i$ and $j$ from $Q_{ij}$. The $R_{ij}$ in the $\dlog$,
\begin{align}
\label{eq:quadric_root_ratio}
 R_{ij} = \frac{Q^{-1}_{ij} + \sqrt{\left(Q^{-1}_{ij}\right)^2 - Q^{-1}_{ii}Q^{-1}_{jj}}}
							{Q^{-1}_{ij} - \sqrt{\left(Q^{-1}_{ij}\right)^2 - Q^{-1}_{ii}Q^{-1}_{jj}}}\,,
\end{align}
is a ratio of the roots of the quadric restricted to rows and columns $i$ and $j$. 

The relation between both representations, \eqref{eq:dgondiffeq} and \eqref{eq:diffMotivic}, can be understood by looking at the completeness relation in embedding space
\begin{equation}
  (YY)= 2 \frac{(YX_+)(Y X_-)}{(X_+X_-)} + \sum\limits^{D}_{a,b=1} Q^{-1}_{ab} (YX_a)(YX_b)\,,
\end{equation}
which is explained in Appendix~\ref{app:embedding}. For the partial fraction analysis of the $D$-gon, we are supposed to isolate all contributions with $D-2$ propagators where the singularities of the $\dlog$-form intersect with the real integration region. This information is extracted from setting the $D-2$ propagators as well as the special ``propagators'' involving the leading singularity points $(YX_+)$ and $(YX_-)$ to zero. The resulting completeness relation generically leads to homogeneous quadratic equations for the remaining propagators evaluated at the special points
\begin{equation}
  0 = Q^{-1}_{ii} (X_i X^{ij}_{\pm})^2  + 2\, Q^{-1}_{ij} (X_i X^{ij}_{\pm})(X_j X^{ij}_{\pm}) + Q^{-1}_{jj} (X_j X^{ij}_{\pm})^2\,,
\end{equation}
which can be easily solved for the ratios $(X_i X^{ij}_{\pm})/(X_j X^{ij}_{\pm})$ which correspond to the two roots, thus exactly reproducing  (\ref{eq:quadric_root_ratio})
\begin{equation}
 R_{ij} =  \frac{(X_iX^{ij}_+)}{(X_jX^{ij}_+)}\frac{(X_jX^{ij}_-)}{(X_iX^{ij}_-)} = u_{ij}\,.
\end{equation}
Our $\dlog$ differential equations are valid for both even and odd spacetime dimensions. The only difference is that the final integration step for odd-dimensional integrals lands us on tadpole integrals in $D=1$ which have transcendental weight zero (if normalized by appropriate powers of $\pi$)
\begin{equation}
I_1 = \int\limits_{-\infty}^{\infty} \frac{d\ell}{\pi}\frac{m}{\ell^2+m^2} = 1\,.
\end{equation} 
Similar differential equations have also been derived in the context of the `graphical co-action' for dimensionally regularized Feynman integrals in \cite{Abreu:2017enx}. There, it was also pointed out that dimension shifted integrals provide a natural one-loop basis.

\subsection{Parity-odd $(D+1)$-gons in $D$ dimensions}

At one loop, there is a special class of integrals that integrates to zero due to spacetime parity, 
where a parity-odd integrand is integrated over the parity-even contour $\Sigma_D$. 
Probably the most well known such integrals is the parity-odd pentagon integral in $D=4$ 
that appears in the standard Passarino-Veltman reduction of one-loop scattering 
amplitudes \cite{Passarino:1978jh}. Here, we briefly explain how this class of 
integrals gives zero from the $\dlog$ perspective as well. 

In embedding space, the general parity-odd $(D+1)$-gons in $D$ dimensions can be written as,
\begin{equation}
\label{eq:dp1_gon_d_dim_rat}
  I^{(D)}_{D+1}[\varepsilon] = \int\limits_{\Sigma_D} \frac{ \langle Y dY^{D+1}\rangle \, \langle YX_1X_2\ldots X_{D+1}\rangle}{(YX_1)(YX_2)\cdots(YX_D)(YX_{D+1})}\,.
\end{equation}
where $\langle \ldots\rangle$ denotes the skew-symmetric contraction of vectors. From the above form 
it is easy to see that this integral evaluates to zero due to Lorentz invariance. The final result has to be proportional 
to the skew symmetric tensor with $D{+}2$ slots, but there are only $D{+}1$ independent vectors available in the problem. 

How can we see this property from the $\dlog$ perspective? The $\dlog$ form of (\ref{eq:dp1_gon_d_dim_rat}) is given by
\begin{equation}
\label{eq:parity_odd_dlog}
I^{(D)}_{D+1}[\varepsilon] = \int\limits_{\Sigma_D} \dlog \frac{(YX_1)}{(YX_2)}\dlog\frac{(YX_2)}{(YX_3)} \cdots  \dlog \frac{(YX_{D})}{(YX_{D+1})}\,.
\end{equation}
This $\dlog$ form, as opposed to the ones in the previous section, does not include 
any additional singularities beyond the propagators. Thus, in the localization procedure the are no 
singularities intersecting the integration cycle $\Sigma_D$ that need to be excised. One therefore 
finds that the integrand is a genuine total derivative and the integral equals zero by Stokes theorem.

\section{Residue theorems and the minimal symbol alphabet}
\label{sec:rt}
%
In the previous section, we have explained that there exists a hierarchy of Feynman integrals in different dimensions related by the differential equation
\begin{align}
\label{eq:hierarchy}
dI^{(D)} =  \frac12 \sum_{i<j} \dlog{u_{ij}}\,I_{ij}^{(D-2)}\,.
\end{align}
The experienced reader might observe that the number of terms in Eq.\eqref{eq:hierarchy} does not depend on the number of scales in the integral. For instance for the $n$-gon integral in $D\!=\!n$ we found $\binom{n}{2}$ $\dlog$s, and hence the same number of possible last entries in its symbol. However, integrals with fewer scales should have fewer letters in their symbol alphabet. From this we might suspect that there exist a yet unknown set of relations between the letters, $u_{ij}$. In this section we explain that the localization procedure contains more information than the final differential equation \eqref{eq:hierarchy}, and how this information provides the missing relations.

The principal observation is that all the objects on the right hand side of the equation arise from a single integrand
\begin{equation}
  dI^{(D)} =  \int\limits_{\Sigma_{D-2}} \Omega^{(D-2,1)}\,.
\end{equation}
The superscript $(r_i,r_e)$, as in section \ref{sec:toymodels}, denotes the degree of the form in loop and external variables respectively.
The singularities of $\Omega^{(D-2,1)}$ are manifest in its $\dlog$ form, but they might be obscured by our choice of integrand basis for generalized unitarity 
\begin{equation}
  I_{ij}^{(D-2)} = \int\limits_{\Sigma_{D-2}} \Omega^{(D-2,0)}_{ij}\,,
\end{equation}
since the $\Omega^{(D-2,0)}_{ij}$ might have spurious singularities which are absent in $\Omega^{(D-2,1)}$.
The residue theorems arise from imposing the cancellation of such spurious singularities\footnote{In two- and four dimensions these singularities correspond to the soft-collinear regions of the loop momentum. For a discussion of residue theorems and their relation to the IR structure of one-loop Feynman parameter integrals, see the recent work in Ref.~\cite{YelleshpurSrikant:2019khx}.}. More concretely, if a subset of the forms $\Omega_{ij}^{(D-2,0)}$ share spurious singularities at the maximal codimension variety $S$ one obtains the relation\footnote{The right hand side of the multiplicative relation is in general a number not equal but the difference is immaterial in the differential equation.}
\begin{equation}
  0= {\rm Res}_S(\Omega^{(D-2)})  = \frac12 \sum_{i<j} a_{ij} \, \dlog u_{ij} \quad \longleftrightarrow \quad  \prod_{i<j} \left(u_{ij}\right)^{a_{ij}} = 1
\end{equation}
where $a_{ij} = \pm1$ or $0$ since the $\Omega_{ij}^{(D-2,0)}$ are $\dlog$ forms and have unit leading singularities. As promised the residue theorem provides multiplicative relations between the different letters that reduce the size of the alphabet. In the following we will illustrate these ideas through a particular example.

\subsubsection*{Massless hexagon in $D=6$}
%
Consider the massless hexagon integral in $D=6$. 
\begin{equation}
I^{(6)}_6= \raisebox{-35pt}{\includegraphics[scale=0.5, trim={0cm 0cm 0cm 0cm},clip]{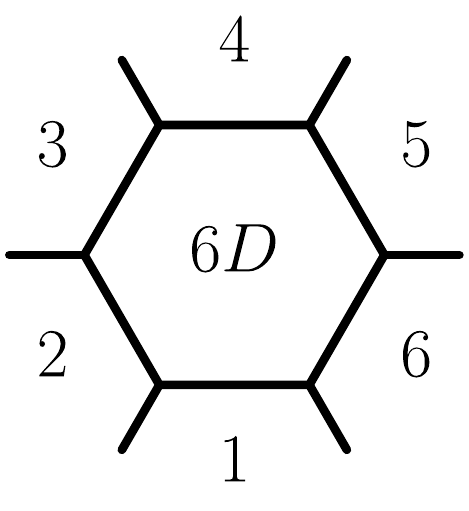}}
\label{eq:hexagon6d}
\end{equation}
This integral was originally calculated in \cite{Dixon:2011ng, DelDuca:2011ne}. It is IR- and UV-finite and naively depends on nine Mandelstam invariants, or eight dimensionless variables.
However, it is well known that it enjoys dual conformal symmetry which implies a restricted kinematic dependence on three cross ratios, commonly called $u,v$ and $w$. Consequently \eqref{eq:hexagon6d} belongs to the space of hexagon-functions, see e.g.~\cite{Dixon:2011pw,Dixon:2013eka}.
Its $\dlog$ integrand and differential equation have the general forms \eqref{eq:ngon_n_dimensions_dlog} and \eqref{eq:dgondiffeq} respectively. In particular its differential equation is written as a sum over fifteen different one- and two-mass boxes in $D=4$,
\begin{equation}
  d I_6^{(6)} =  \sum_{i< j} \dlog{u_{ij}} \, I_{4,\,ij}^{(4)}\,.
\end{equation}
all of which are IR divergent. In four dimensions IR singularities generally arise from loop-integration regions where the loop momenta becomes collinear to one of the massless external momenta, or soft \cite{ArkaniHamed:2010gh,Bourjaily:2013mma}.  These divergences manifest themselves at the \emph{integrand} level as the possibility of accessing certain \emph{composite} residues where not only propagators but also Jacobians are set on shell. In our situation, the composite poles arise from the Jacobians of triple cuts with either one or two massless corners. On the support of such cuts the loop momentum lies in the region that gives rise to the divergences of the integral.
\begin{figure}[t]
  \centering
  collinear: \hspace{5pt} \raisebox{-28pt}{\includegraphics[scale=0.42]{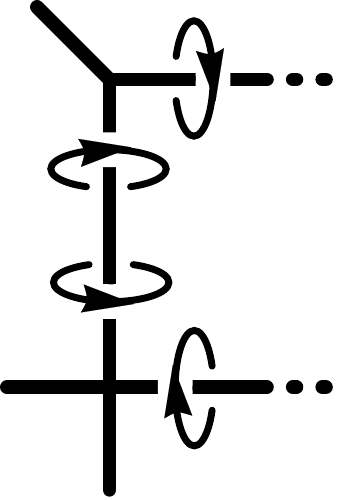}}
  \hspace{1.2cm}
  soft-collinear: \hspace{10pt} \raisebox{-25pt}{\includegraphics[scale=0.4]{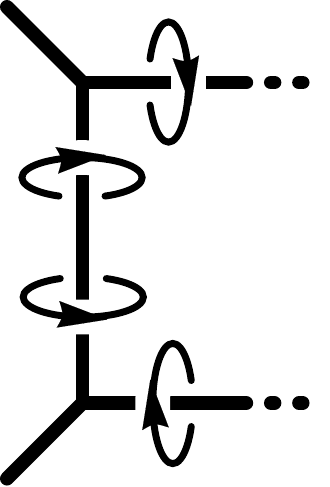}} 
  \caption{\label{fig:ir_cuts}Spurious IR singularities from individual boxes. The double circle indicates a composite residue of the three propagators as well as the Jacobian.}
\end{figure}
Both of these singularity are absent in $\Omega^{(4,1)}_6$ (since $I^{(6)}_6$ is IR finite). The absence of the soft and collinear composite residues at the level of the \emph{integrand} $\Omega^{(4,1)}_6$, implies a number of nontrivial relations between the coefficients of $\Omega^{(4,0)}_{ij}$. All relevant relations correspond to residue theorems involving three cut propagators and a subsequent soft- or collinear cut of the Jacobian, as illustrated in figure \ref{fig:ir_cuts}. A concrete example is given by cutting the three propagators $(Y4)=(Y5)=(Y6)=0$, which results in a function 
\begin{align}
  \Omega^{(4,1)}_6\big|_\text{cut}(z) =
      \dlog{u_{12}}  \includegraphics[scale=0.4, trim= -10 77 0 0]{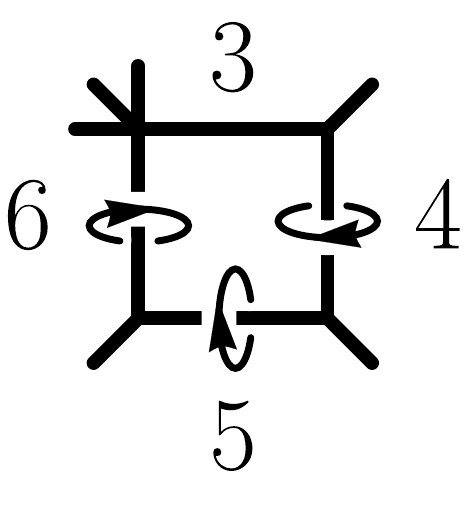}
     + \dlog{u_{23}} \includegraphics[scale=0.4, trim= -10 77 0 0]{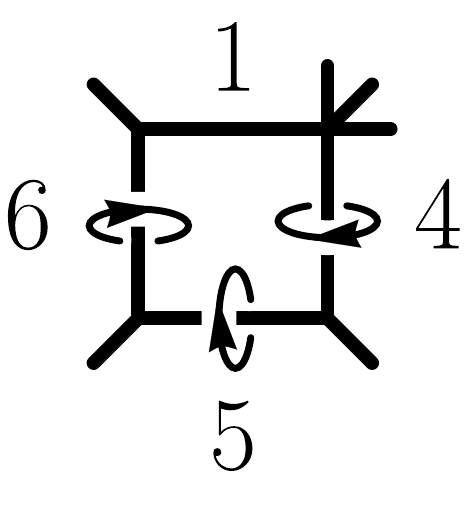}
     + \dlog{u_{13}} \includegraphics[scale=0.4, trim= -10 77 0 0]{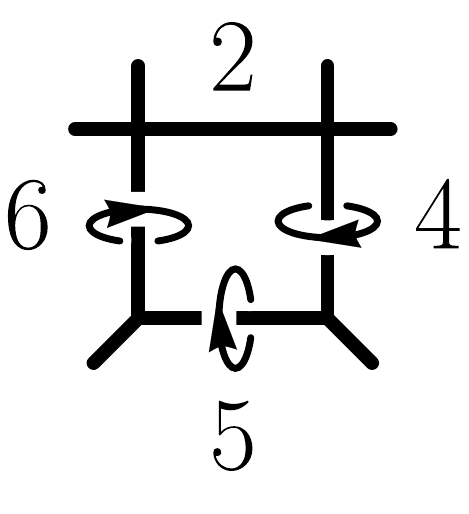} \,.
  \label{eq:hex3cut}
     \\ \nonumber
\end{align}
of the single remaining degree of freedom $z$\footnote{In a familiar parameterization of the loop momentum in spinor-helicity variables this is given by $\ell = z \lambda_4 \tilde{ \lambda}_5$ or its parity conjugate}. Every one of the contributing boxes has a spurious pole in the Jacobian, $\J(z)$, of the triple cut. Requiring the absence of such a singularity in $\Omega^{(4,1)}_6$ gives the following relation between the individual pieces
\begin{equation}
  0= \text{Res} \big[\Omega^{(4)}_6\big|_\text{cut}(z),\J(z)=0\big] =  \dlog{u_{12}}   + \dlog{u_{23}}  + \dlog{u_{13}} = \dlog{u_{12}u_{23}u_{13}}\,,
   \label{eq:hexresidueth}
 \end{equation}
which can be solved as $u_{13}= (u_{12}u_{23})^{-1}$. There exist five additional relations obtained from triple cuts involving two massless corners as well as 12 more relations coming from the collinear regions of a single massless corner. Only 12 of these 18 relations are linearly independent allowing us to reduce the naive 15 final entries to  only three independent letters, which we can choose to be $u_{12}, u_{34}$ and $u_{56}$. In terms of this minimal set of final entries, the differential equation takes its most compact form
\begin{align}
  \begin{split} 
  d \, \raisebox{-34pt}{\includegraphics[scale=0.5]{figures/hex.pdf}} &= \dlog u_{12} \left(  
  															  \raisebox{-34pt}{\includegraphics[scale=0.6]{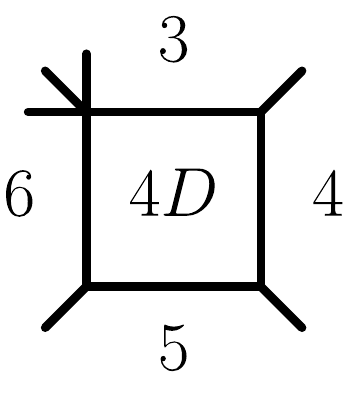} }
  															- \raisebox{-34pt}{\includegraphics[scale=0.6]{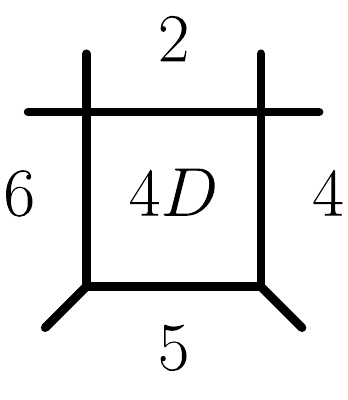}} 
															+ \raisebox{-34pt}{\includegraphics[scale=0.6]{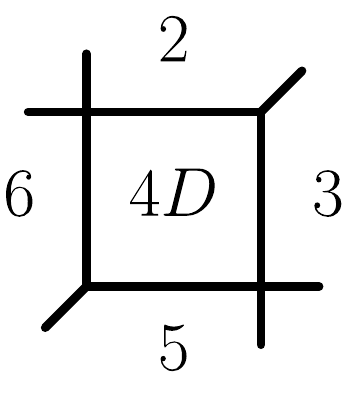}} \right. \\[-5pt]
  & \hspace{55pt} + \raisebox{-34pt}{\includegraphics[scale=0.6]{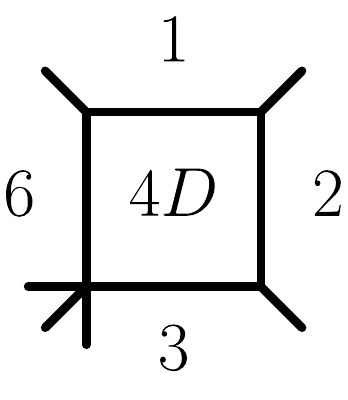}} 
  			   - \raisebox{-34pt}{\includegraphics[scale=0.6]{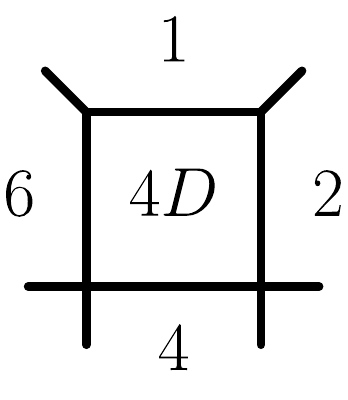}}
			   +\raisebox{-34pt}{\includegraphics[scale=0.6]{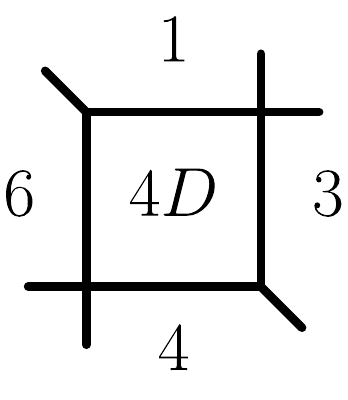}}  \\[-5pt]
  & \left. \hspace{55pt}+ \hspace{2pt} 
  				\raisebox{-34pt}{\includegraphics[scale=0.6]{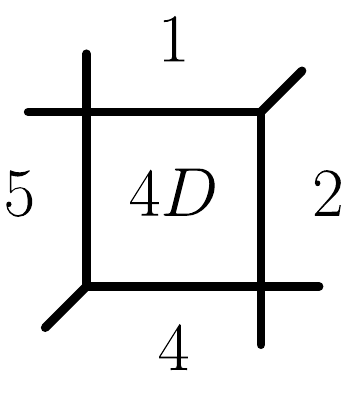}} 
			      - \raisebox{-34pt}{\includegraphics[scale=0.6]{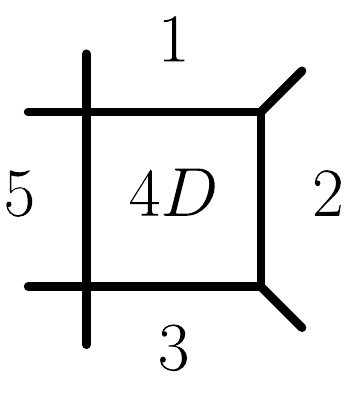}} 
			      - \raisebox{-34pt}{\includegraphics[scale=0.6]{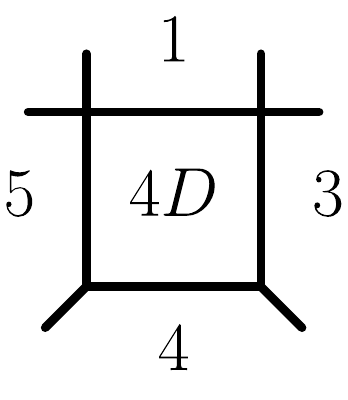}} \right) \\[0pt]
			      & \hspace{55pt} + \quad (345612) \quad + \quad (561234) \,,
  \hspace{-2cm}
  \end{split} 
\end{align}
where $(345612)$ and $(561234)$ denote the two cyclic permutations of $(123456)$ by two and four. As required, in this representation the three sums of boxes accompanying each $\dlog u_k$ are IR finite.
Each particular combination of boxes might at first seem arbitrary, but they turn out to be exactly the parity-even parts of the IR finite hexagons with special chiral numerator defined in \cite{ArkaniHamed:2010gh}
\vspace{-.3cm}
\begin{equation}
 \raisebox{-34pt}{\includegraphics[scale=0.5]{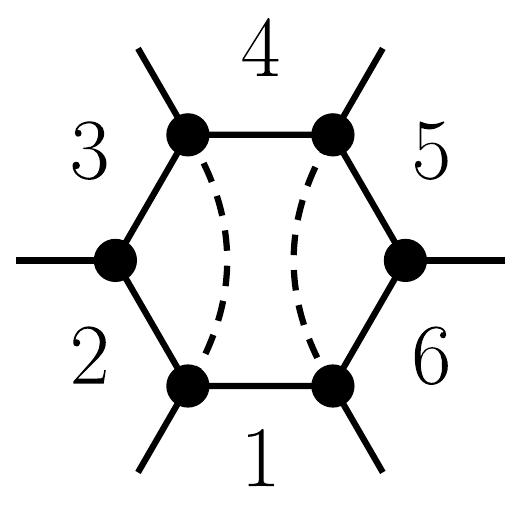} } 
 	=  \int\limits_{\Sigma_4} \frac{\langle Y d^5Y\rangle \, (X_1X_4)(X_2X_5)(X_3X_6) (YX_+^{23}) (YX_+^{56})}{(X_+^{56}X_+^{23}) (YX_1)(YX_2)(YX_3)(YX_4)(YX_5)(YX_6)}\,,
\vspace{-.1cm}
\end{equation}
In addition, the three independent letters correspond to the parity-odd $y$ variables in the hexagon alphabet (see e.g. \cite{Dixon:2013eka})
\begin{equation}
  u_{12} = y_v\,, \quad u_{34} = y_u\,, \quad u_{56} = y_w \,.
\end{equation}
With these definitions, we reproduce the known result of \cite{Dixon:2011ng}
\begin{equation*}
  dI^{(6)}_6 =  d\log (y_w) \, \raisebox{-34pt}{\includegraphics[scale=0.5]{figures/1loop_Omega_uvw.pdf}}
            + d\log (y_v) \, \raisebox{-34pt}{\includegraphics[scale=0.5]{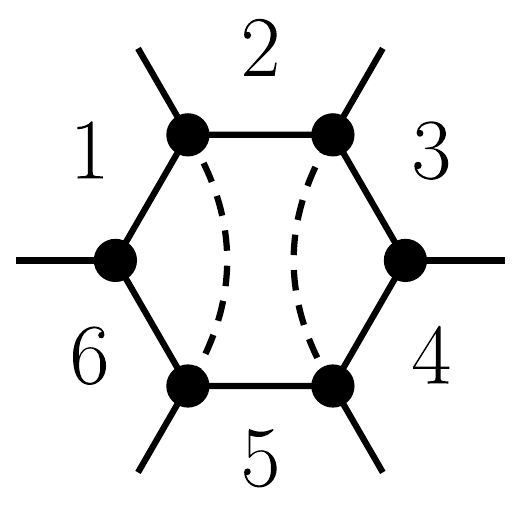}} 
            + d\log (y_u) \, \raisebox{-34pt}{\includegraphics[scale=0.5]{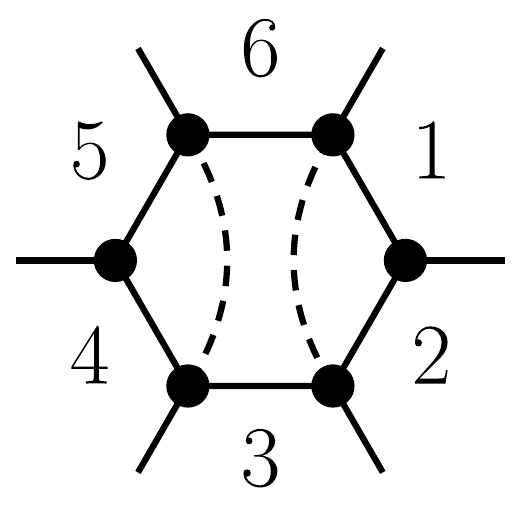}} \,.
\end{equation*}
Similar relations to the ones we have presented here were observed at the amplitude level between the coefficients of different integrals in one-loop gauge theory amplitudes. These go by the name of ``IR-equations'' and follow from the well-known IR factorization properties of loop amplitudes and the cancellation of spurious singularities \cite{Giele:1991vf,Kunszt:1994np}. These have been reinterpreted as following from residue theorems at the integrand level in Sec.~3.5 of Ref.~\cite{Bourjaily:2013mma}. In this work have applied a similar reasoning to individual loop-integrals rather than full amplitudes.

%
\section{Higher-loop examples}
\label{sec:twoloop}
%

\subsection{Toy integral}
\label{subsec:twolooptoy}

Before dealing with the complexities of an actual Feynman integral, we would like to introduce the relevant features on a very simple one-dimensional toy integral
\begin{equation}
 I_t = \int\limits_0^1 \dlog(x+a) \log\left( \frac{x+b}{c}\right)\,.
\end{equation}
An algorithm for evaluating single-variable integrals of the form $I = \int \dlog f(x) \times G(x)$, where $G$ is some transcendental function has been described in Appendix A of \cite{CaronHuot:2011kk}, see also \cite{Panzer:2015ida,Henn:2018cdp}.  As before, instead of evaluating this integral directly, we take a differential with respect to the external parameters $(a,b,c)$ and simply use the Leibniz rule. There are two kinds of contributions: 
\begin{equation}
d I_t = I^{\text{bdry}}_t + I^{\text{bulk}}_t\,.
\end{equation}
The first, $I^{\text{bdry}}_t$ comes from the boundary term
\begin{equation}
 I^{\text{bdry}}_t  = \dlog(x+a) \log\left( \frac{x+b}{c}\right) \bigg|^1_0 = \dlog(a+1)   \log\left( \frac{b+1}{c}\right) - \dlog a \,  \log\left( \frac{b}{c}\right)\,.
 \label{eq:ItoyBdy}
\end{equation} 
as in the single variable example of Eq.\eqref{eq:SingleVarEx}.
The second term $I^{\text{bulk}}_t$ is given by 
\begin{equation}
I^{\text{bulk}}_t = \int\limits_0^1 \dlog(x+a) \dlog\left( \frac{x+b}{c}\right)\,,
\end{equation}
which we need to partial fraction in order to extract a $\dlog$ of the external variables. In this one variable case, partial fractioning the integrand is a trivial operation, yielding
\begin{equation}
 \dlog(x+a) \dlog\left( \frac{x+b}{c}\right) = \dlog(a-b)  \dlog \left(\frac{x+a}{x+b}\right) - \dlog c\, \dlog(x+a)\,.
\end{equation}
Using this form of the integrand, we can evaluate $I^{\text{bulk}}_t$
\begin{equation}
 I^{\text{bulk}}_t = \dlog(a-b)  \log \left(\frac{(1+a)}{(1+b)} \frac{b}{a}\right)  - \dlog c\, \log\left(\frac{1+a}{a}\right)\,.
 \label{eq:ItoyBulk}
\end{equation}
Combining Eqs.~\eqref{eq:ItoyBdy} and \eqref{eq:ItoyBulk} gives the full differential of $dI_t$.

We will see momentarily that the extra ingredient of this toy example, relative to the one in section \ref{subsec:onevtoymodels}, is precisely what  is needed to study the differential of two- and higher-loop $\dlog$ Feynman integrals.

\subsection{Two-loop off-shell ladder}
\label{subsec:twoloopladder}

As a proof of concept, and to show that the differential equations for $\dlog$ integrals are not limited to one-loop integrals,  we now discuss one concrete two-loop example: the two-loop off-shell ladder
\begin{align}
I^{2\text{-loop}}_{4m} & = \raisebox{-35pt}{
\includegraphics[scale=.5]{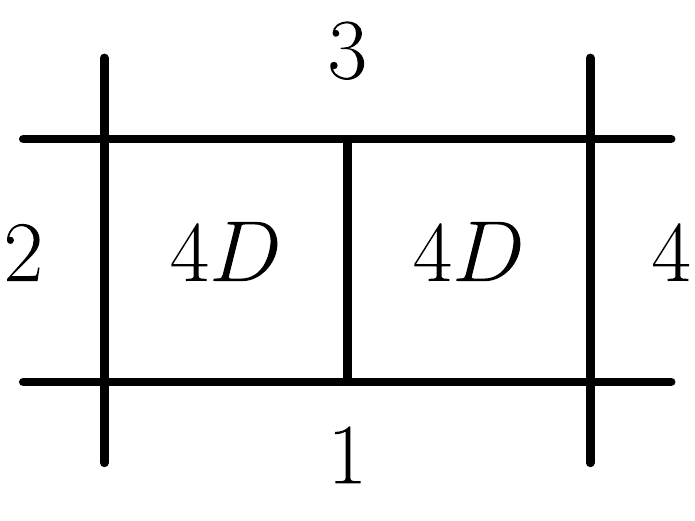}}\,.
\end{align}
In embedding space formalism, this integral is given by
\begin{align}
  I^{2\text{-loop}}_{4m} & = \int\limits_{\Sigma_4^L \cup \Sigma_4^R} \frac{\langle Y_L d^5 Y_L \rangle \,\langle Y_R d^5 Y_R \rangle \ (X_1X_3) \sqrt{-\det(X_iX_j)}}{(Y_LX_1)(Y_LX_2)(Y_LX_3)(Y_LY_R)(Y_RX_3)(Y_RX_4)(Y_RX_1)}\,.
 \label{eq:2loop_4ml_double_box_rat_form}
\end{align}
This integral is of course well known \cite{Usyukina:1993ch,Broadhurst:2010ds} and has a number of special properties. It is UV and IR finite as well as dual conformally invariant which restricts the kinematic dependence to two dual conformal cross-ratios,
\begin{equation}
\label{eq:xratdef}
 u = z\zb = \frac{(X_1X_2)(X_3X_4)}{(X_1X_3)(X_2X_4)}\,, \quad 
 v = (1-z)(1-\zb) =  \frac{(X_2X_3)(X_4X_1)}{(X_1X_3)(X_2X_4)}\,.
\end{equation}

As a first step for deriving the differential equation for this integral, we integrate out the right-hand-side box, $I_{4,R}^{(4)}\equiv I_{4,R}^{(4)}(X_1,Y_L,X_3,X_4)$, 
\begin{align}
  I^{2\text{-loop}}_{4m} & = \int\limits_{\Sigma_4^L} \frac{\langle Y_L d^5 Y_L \rangle \, \ (X_1X_3) \sqrt{-\det(X_iX_j)}}{(Y_LX_1)(Y_LX_2)(Y_LX_3) \sqrt{\Delta}} I_{4,R}^{(4)} \equiv \int\limits_{\Sigma_4^L} \omega_L I_{4,R}^{(4)} \,,
  \label{eq:2loopintegratedout}
\end{align}
and end up with a one-loop form, $\omega_L$, with three propagator poles and the square-root of the Gram determinant of the right box $\Delta = -\det(X_1,Y_L,X_3,X_4)$. We call this object, which was first recognized in Ref.~\cite{Buchbinder:2005wp}, a \emph{generalized box}. Note that the apparent singularity $\sqrt{\Delta}\rightarrow 0$ is absent in Eq.\eqref{eq:genboxdlog}, since $I_{4,R}^{(4)}$ vanishes in this limit. We find a $\dlog$ form for the generalized box
\begin{align}
  \omega_L =   \frac12 \dlog\left(\frac{Y_LX_1}{Y_L X_2}\right)\dlog\left(\frac{Y_LX_2}{Y_L X_3}\right)\dlog\frac{(X_1X^{13}_{+,R})(X_3X^{13}_{-,R})}{(X_1X^{13}_{-,R})(X_3X^{13}_{+,R})}\dlog\left(\frac{Y_LX_+^L}{Y_L X_-^L}\right)\,,
  \label{eq:genboxdlog}
\end{align}
where 
\begin{align}
  \begin{split}
  X_{\pm}^L \quad \text{are the solutions to} \quad &(Y_LX_1) = (Y_LX_2) = (Y_LX_3) = (Y_LX_4) =0 \,,\\
  X_{\pm}^R \quad \text{are the solutions to} \quad &(Y_RX_1) = (Y_RY_L) = (Y_RX_3) = (Y_RX_4) =0\,,\\
  X^{13}_{\pm,R} \quad \text{are the solutions to} \quad &(Y_RX_+^R) = (Y_RX_-^R) = (Y_RY_L) = (Y_RX_4) =0\,.
  \label{eq:x13pmRdef}
\end{split} 
\end{align}
It is worth noting that $X_{\pm}^L$ are the cut solutions to the would-be left hand box, with missing propagator $(Y_LX_4)$. Similarly, $X_{\pm}^R$ are the solutions to the right box cut, and $X^{13}_{\pm,R}$ are the points featuring in our formula for its differential equation derived in Section~\ref{sec:oneloop}. Note also that $X_{\pm}^R$ and $X^{13}_{\pm,R}$ depend implicitly on the unintegrated variable $Y_L$. The unfamiliar argument in the third entry of the $\dlog$ in Eq.\eqref{eq:genboxdlog} is actually a last entry of the symbol of $I_{4,R}^{(4)}$ as given by the general one-loop formula \eqref{eq:dgondiffeq}. The fact that $\sqrt{\Delta}$ does not introduce an additional singularity in the integral suggests that the third entry of the $\dlog$ should contain factors of the schematic structure $a\pm\sqrt{\Delta}$. Indeed, 
\begin{equation}
  \frac{(X_1X^{13}_{+,R})(X_3X^{13}_{-,R})}{(X_1X^{13}_{-,R})(X_3X^{13}_{+,R})} = \frac{a+\sqrt{\Delta}}{a-\sqrt{\Delta}}\,,
  \label{eq:apmdelta}
\end{equation}
where
\begin{equation}
  a = (X_1X_3)(Y_LX_4) - (X_1X_4)(Y_LX_3) - (X_3X_4)(Y_LX_1)\,.
\end{equation}
The kind of argument in Eq.\eqref{eq:apmdelta} only becomes singular when $a^2-\Delta=0$, but
\begin{equation}
\label{eq:asqmDelta}
  a^2-\Delta = 4(X_1X_4)(X_3X_4)(Y_LX_1)(Y_LX_3)
\end{equation}
so the only singularities are the usual propagators. It is important to keep this relation in mind, since several entries of the $\dlog$ form in Eq.\eqref{eq:genboxdlog} contain identical singularities and one needs to be careful when calculating residues.

\subsubsection*{Localization and generalized unitarity}

With the $\dlog$ form of $\omega_L$ in hand, we now derive the differential equation for this two-loop integral starting from Eq.\eqref{eq:2loopintegratedout}. Essentially, what we utilize here amounts to a generalization of the single-variable integration algorithm described above. As always we start by taking derivatives with respect to the external variables and rewrite them in terms of a total derivative and additional pieces
\begin{equation}
d_e I^{2\text{-loop}}_{4m}  = \int\limits_{\Sigma_4^L} \left[d_e\omega^{(4,0)}_L\, I_{4,R}^{(4)} + \omega_L^{(4,0)}\,d_eI_{4,R}^{(4)}\right] = 
	\int\limits_{\Sigma_4^L} \left[-d_i\left(\omega^{(3,1)}_L\, I_{4,R}^{(4)}\right) + 
	\omega_L \,dI_{4,R}^{(4)} \right]
\vspace{-0.2cm}
\end{equation}
Unlike the one-loop case, here we find two kinds of contributions.
\begin{equation}
\label{eq:f1_def_2loop}
  F_1=-\int\limits_{\Sigma_4^L} d_i\left(\omega^{(3,1)}_L\, I_{4,R}^{(4)}\right)\,, \quad F_2=\int\limits_{\Sigma_4^L}  \omega_L \,dI_{4,R}^{(4)}\,.
\vspace{-0.2cm}
\end{equation}
The first contribution, $F_1$, is a total derivative which results in the localization of the generalized box on the left hand side, just as at one-loop. This is the analog of the boundary term in the algorithm of \cite{CaronHuot:2011kk}. Starting at two loops, the second term, $F_2$, is new and arises from the differential acting on the integrated box on the right. This piece also localizes by the general formula in Eq.\eqref{eq:dgondiffeq}.

\medskip 

Let us start by analyzing $F_2$. Somewhat surprisingly, we find that there is actually no contribution coming from this term, i.e. $F_2 =0$. Let us briefly explain why. Making use of the one-loop result obtained in subsec.~\ref{subsec:dgonddim} and the relations of Section \ref{sec:rt}, we see that the differential of the box integral, $dI_{4,R}^{(4)}$, only has two independent final entries. Therefore one is left with two different five-forms to partial fraction,
\begin{align}
 \omega_L \ \dlog\frac{(X_1X^{1Y_L}_{+,R})(Y_LX^{1Y_L}_{-,R})}{(X_1X^{1Y_L}_{-,R})(Y_LX^{1Y_L}_{+,R})}\,,
 \qquad
 \omega_L \ \dlog\frac{(X_3X^{3Y_L}_{+,R})(Y_LX^{3Y_L}_{-,R})}{(X_3X^{3Y_L}_{-,R})(Y_LX^{3Y_L}_{+,R})}
\end{align}
One can explicitly check that both forms are zero, either by writing them as rational forms\footnote{Converting the $\dlog$-form with entries $f_j (x_i)$ to a rational form in the $x_i$ involves computing the Jacobian $\det\! \left(\!\frac{\partial \log f_j}{\partial x_i}\!\right)$. In the case discussed here, the Jacobian vanishes implying a linear relation between the $\dlog$ factors with $x_i$ dependent coefficients. Importantly, this  \emph{does not} necessarily imply that there is a multiplicative relation between the $f_j$. } or by checking that all residues are zero. At this point, it is unclear whether or not this happens for more general integrals. In any case, one could partial fraction the resulting form using generalized unitarity and derive the corresponding contribution to the differential equation. 

Let us now study $F_1$ in detail. As explained above, the only additional singularities in the $\dlog$ form in Eq.\eqref{eq:genboxdlog} are the by now familiar $(Y_LX^L_\pm)$ in the last slot. Thus the fate of the total derivative is exactly the same as at one loop, one has to excise the corresponding singularities at $\Sigma_2^L = \Sigma_4^L\cap  \{(Y_LX^L_\pm) =0\}$ from the integration cycle, which produces a boundary term by Stokes theorem. As before, one further integration is done by a residue computation, which yields
\begin{equation}
  F_1 = -\int\limits_{\Sigma_2^L} \Omega^{(2,1)}_L\, I_{4,R}^{(4)} \,,
  \label{eq:F1_localized}
  \vspace{-0.4cm}
\end{equation}
where
\begin{equation}
 \Omega^{(2,1)}_L = \text{Res}_{\Sigma_2^L}[\omega^{(3,1)}_L] =   \dlog\left(\frac{Y_LX_1}{Y_L X_2}\right)\dlog\left(\frac{Y_LX_2}{Y_L X_3}\right)\dlog\frac{a+\sqrt{\Delta}}{a-\sqrt{\Delta}} \,.
 \label{eq:Omega21Ldef}
\end{equation}
As in the one-loop examples, we proceed to partial fraction $\Omega^{(2,1)}_L$ in Eq.\eqref{eq:Omega21Ldef} in order to pull out one differential that only depends on external kinematic variables. Using generalized unitarity, we can write an ansatz for the two-dimensional integral whose coefficients are fixed by comparing residues. It is clear from Eq.\eqref{eq:Omega21Ldef} that there are only three propagator poles present in the $\dlog$ form, so that we need to obtain the coefficients of three parity-even bubbles, $\Omega^{(2,0)}_{ij}$,  and a parity-odd triangle, $\Omega^{(2,0)}_{123}$
\begin{align}
 \Omega^{(2,1)}_L = c_{12} \Omega^{(2,0)}_{12} + c_{13} \Omega^{(2,0)}_{13} + c_{23} \Omega^{(2,0)}_{23} + c_{123} \Omega^{(2,0)}_{123}
\end{align}
Unlike in our one-loop discussion, one cannot simply drop the parity-odd terms, since they appear in combination with the nontrivial function $I_{4,R}^{(4)}$ in Eq.(\ref{eq:f1_def_2loop}) under the integral sign (see discussion below for more details).

As a concrete example, let us briefly explain how to compute the appropriate residues of $\Omega^{(2,1)}_L$. Our goal is to determine the coefficients of the $\Omega^{(2,0)}_{12}$ bubble integral and the triangle $\Omega^{(2,0)}_{123}$. Using the $\dlog$ identities of Eq.\eqref{eq:dlogidentities}, we rewrite $\Omega^{(2,1)}_L$
\begin{equation}
\label{eq:omega21def}
\Omega^{(2,1)}_L = \dlog\left(\frac{Y_LX_1}{Y_L X_2}\right)\dlog\left(\frac{Y_LX_1}{Y_L X_3}\right)\dlog\frac{a+\sqrt{\Delta}}{a-\sqrt{\Delta}}\,,
\end{equation}
so that only the first slot of the $\dlog$ contains $(Y_LX_2)$ where it is now trivial to take the residue,
\begin{equation}
\underset{(Y_LX_2)=0}{\text{Res}} \left[\Omega^{(2,1)}_L\right] = - \dlog\left(\frac{Y_LX_1}{Y_L X_3}\right)\dlog\frac{a+\sqrt{\Delta}}{a-\sqrt{\Delta}}\,.
\end{equation}
In order to take the second residue in $(Y_LX_1)=0$, we have to remember that the second $\dlog$ has also a singularity at this location and we have to be a bit more careful. To expose this singularity, we make use of Eq.\eqref{eq:asqmDelta} and remove it by using once again, the $\dlog$ identities of Eq.\eqref{eq:dlogidentities}
\begin{equation}
\underset{(Y_LX_2)=0}{\text{Res}} \left[\Omega^{(2,1)}_L\right] = - \dlog\left(\frac{Y_LX_1}{Y_L X_3}\right)\dlog\frac{\left(a+\sqrt{\Delta}\right)^2}{(X_1X_4)(X_3X_4)(Y_LX_3)^2}\,.
\end{equation}
Finally, we can take the residue in $(Y_LX_1)=0$ for which there are two solutions, denoted as $X^{3}_{\pm,L}$ which is given by evaluating the last $\dlog$ on either of these two points.  Then the coefficient of the scalar $\Omega^{(2,0)}_{12}$ bubble is given by
\begin{align}
\label{eq:c12_def}
\hspace{-.9cm}
c_{12}	\! =\! \frac12\! \left({\rm Res}[\Omega^{(2,1)}_L\!,\!X_{+,L}^{3}] \!-\! {\rm Res}[\Omega^{(2,1)}_L\!,\! X_{-,L}^{3}]\right) 
		\!=\! \dlog \frac{(X_1X^{13+}_{+,R})(X_3X^{13+}_{-,R})(X^3_{-,L} X_3)}{(X_1X^{13-}_{+,R})(X_3X^{13-}_{-,R})(X^3_{+,L} X_3)}\,, \hspace{-1cm}
\end{align}
where the superscript $\pm$ in the points $X^{13\pm}_{\pm,R}$ indicate that the implicit dependence of $X^{13}_{\pm,R}$ on $Y_L$ (defined in Eq.\eqref{eq:x13pmRdef}) has been substituted by $X^{3}_{\pm,L}$. Admittedly, the notation here is quite heavy. However it stresses that all arguments of the external $\dlog$s can be written as a ratio of inner products of special points tied to the cut geometry of the integral.

Similarly, one can extract the residue of the parity-odd triangle by simply averaging over the two leading singularities instead of taking the difference
\begin{align}
\label{c123_def}
\hspace{-.9cm}
c_{123}	\! =\! \frac12\! \left({\rm Res}[\Omega^{(2,1)}_L\!,\!X_{+,L}^{3}] \!+\! {\rm Res}[\Omega^{(2,1)}_L\!,\! X_{-,L}^{3}]\right) \,,
\end{align}
since the residues of $\Omega^{(2,0)}_{123}$ at the two solutions $X_{+,L}^{3}$ and $X_{-,L}^{3}$ are $(1,1)$ respectively.

At the end of the day, plugging in all special points, we can write the coefficients in terms of the $z,\zb$ parameters defined in Eq.\eqref{eq:xratdef}
\begin{align}
\begin{split}
\label{eq:c12bubCoeff}
c_{12} = -c_{13} = c_{23}	& = \dlog \left(\frac{z(1-\zb)}{\zb(1-z)}\right)\,, \quad
c_{123} 	 = \dlog \left(\frac{z \zb}{(1-z)(1-\zb)}\right)\,. 
\end{split}
\end{align}
Alternatively, we can explicitly parameterize the form $\Omega^{(2,1)}_L$ in embedding space. The loop momentum is expanded as in Eq.\eqref{eq:embeddingLoop} in terms of the points
 \begin{equation}
   X_1 \!=\! \begin{pmatrix} 0 \\ 0 \\ 0 \\ 0 \\ 0 \\ 1 \end{pmatrix} \,
   X_2 \!=\! \begin{pmatrix} \frac12 (z{-}\zb) \\ \frac12 (z{+}\zb) \\ 0 \\ 0 \\ z \zb \\ 1 \end{pmatrix} \,
   X_3 \!=\! \begin{pmatrix} 0 \\ 1 \\ 0 \\ 0 \\ 1 \\ 1 \end{pmatrix} \,
   X_4 \!=\! \begin{pmatrix} 0 \\0 \\0 \\0 \\1 \\ 0 \end{pmatrix} \,
   X_5 \!=\! \begin{pmatrix} 0 \\0 \\{-}\frac12 \\{-}\frac{i}{2} \\0 \\ 0 \end{pmatrix} \,
   X_6 \!=\! \begin{pmatrix} 0 \\0 \\{-}\frac12 \\{+}\frac{i}{2} \\0 \\ 0 \end{pmatrix} \,,
 \end{equation}
for which we have chosen a convenient parametrization. In terms of these variables, the $\dlog$ form in Eq.\eqref{eq:omega21def} can be evaluated and partial fractioned explicitly. 

The mixed-dimensional integrals resulting from the partial fractioning can be identified from the set of their propagators. For instance we find that 
\begin{align}
\int\limits_{\Sigma_2^L} \Omega^{(2,0)}_{13}\, I_{4,R}^{(4)} 
=
\raisebox{-28pt}{\includegraphics[scale=0.40]{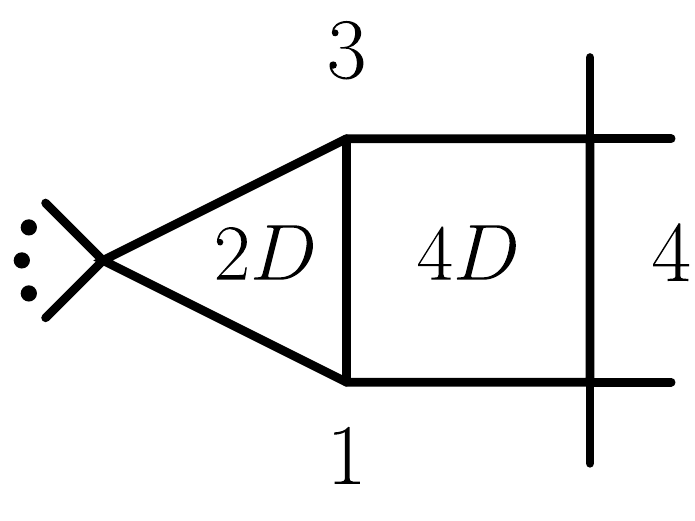}}\,,
\qquad 
\int\limits_{\Sigma_2^L} \Omega^{(2,0)}_{123}\, I_{4,R}^{(4)} 
=
\raisebox{-28pt}{\includegraphics[scale=0.40]{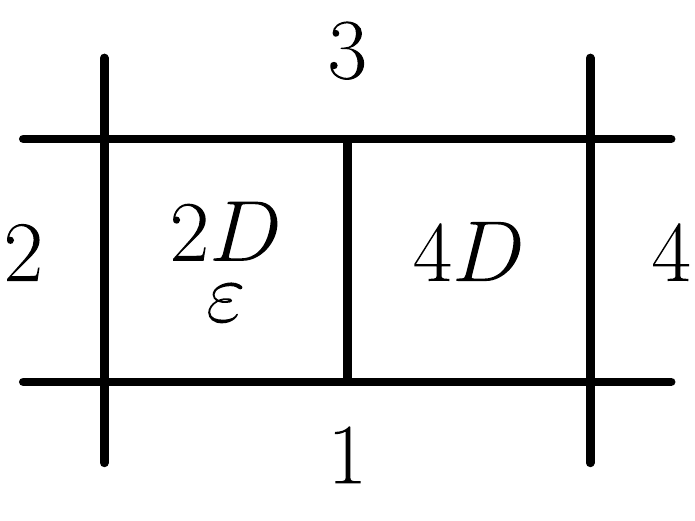}} \,,
\end{align}
where $\varepsilon$ denotes the insertion of the two-dimensional parity-odd numerator $\langle Y_LX_1X_2 X_3\rangle$, and the integrand is normalized to be unit leading singularity loop by loop\footnote{Note that despite the picture, this implies that they carry nontrivial numerators.}. Alternatively, one can ``integrate in'' the right-hand box $I_{4,R}^{(4)}$ and check that the $\dlog$ form of the full integrand yields the rational form corresponding to such integrals. In the two examples above we can write the integrals explicitly as
\begin{align}
  \begin{split}
    &\hspace{-0.4cm}\raisebox{-28pt}{\includegraphics[scale=0.40]{figures/boxtri13.pdf}} =  \int\limits_{\Sigma_2^L\times \Sigma_4^R} 
  \dlog\left(\frac{Y_LX_1}{Y_LX_3}\right) 
  \dlog\left(\frac{Y_LX^2_{+,L}}{Y_LX^2_{-,L}}\right) \\ 
  &\hspace{2cm}\times
  \dlog\left(\frac{Y_RX_1}{Y_RY_L}\right)
  \dlog\left(\frac{Y_RY_L}{Y_RX_3}\right)
  \dlog\left(\frac{Y_RX_3}{Y_RX_4}\right)
  \dlog\left(\frac{Y_RX_{+,R}}{Y_RX_{-,R}}\right) \\
  &\hspace{2.36cm}= \int\limits_{\Sigma_2^L\times \Sigma_4^R} \frac{\langle Y_Ld^3Y_L\rangle \langle Y_Rd^5Y_R\rangle (X_1X_3)\langle X_1 Y_L X_3 X_4  X_{+,R} X_{-,R}\rangle}{(Y_LX_1)(Y_LX_3)(Y_LY_R)(Y_RX_1)(Y_RX_3)(Y_RX_4)(X_{+,R}X_{-,R})}
  \end{split}
\end{align}
\vskip -.4cm
%
\begin{align} 
  \begin{split}
 &\hspace{-0.4cm}\raisebox{-28pt}{\includegraphics[scale=0.40]{figures/double_box_massive_2d4d.pdf}} =  \int\limits_{\Sigma_2^L\times \Sigma_4^R} 
  \dlog\left(\frac{Y_LX_1}{Y_LX_2}\right) 
  \dlog\left(\frac{Y_LX_2}{Y_LX_3}\right) \\ 
  &\hspace{2cm}\times
  \dlog\left(\frac{Y_RX_1}{Y_RY_L}\right)
  \dlog\left(\frac{Y_RY_L}{Y_RX_3}\right)
  \dlog\left(\frac{Y_RX_3}{Y_RX_4}\right)
  \dlog\left(\frac{Y_RX_{+,R}}{Y_RX_{-,R}}\right) \\
  &\hspace{2.36cm}= \int\limits_{\Sigma_2^L\times \Sigma_4^R} \frac{\langle Y_Ld^3Y_L\rangle \langle Y_Rd^5Y_R\rangle \langle Y_L X_1 X_2 X_3\rangle \langle X_1 Y_L X_3 X_4  X_{+,R} X_{-,R}\rangle}{(Y_LX_1)(Y_LX_3)(Y_LY_R)(Y_RX_1)(Y_RX_3)(Y_RX_4)(X_{+,R}X_{-,R})}
  \end{split}
\end{align}
The appearance of such mixed-dimension integrals is quite natural, since one expect the differential of the weight-four ladder to produce weight-three objects. However, it is remarkable that these weight-three objects are also Feynman integrals. It would be very interesting to evaluate these integrals using more conventional methods. Also it will be important to understand if and how they are related to the basis of four-dimensional integrals that feature in the canonical differential equations, similar to the one-loop example in Sec.~\ref{sec:oneloop}.

Putting all the pieces together we find the following result the differential of the two-loop off-shell ladder 
\begin{align}
\hspace{-.5cm}
   d I^{2\text{-loop}}_{4m}  & =   	 \dlog \left(\frac{z(1-\zb)}{\zb(1-z)}\right) \!\! \left[
\raisebox{-28pt}{\includegraphics[scale=0.40]{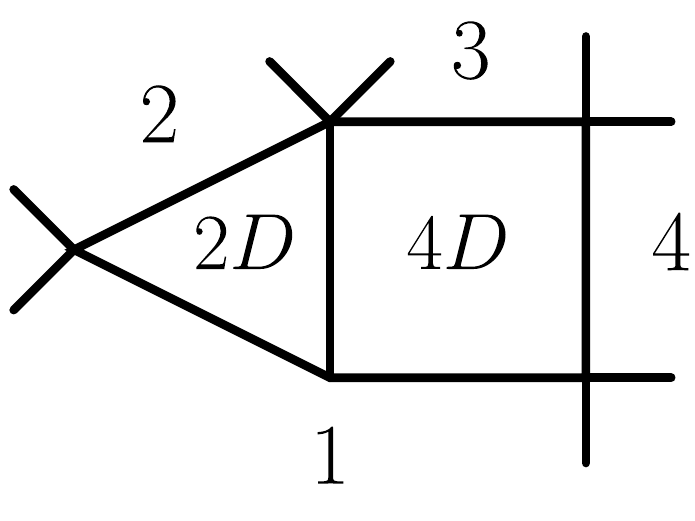}} \;
- 
\raisebox{-28pt}{\includegraphics[scale=0.40]{figures/boxtri13.pdf}} \;
+
\raisebox{-28pt}{\includegraphics[scale=0.40]{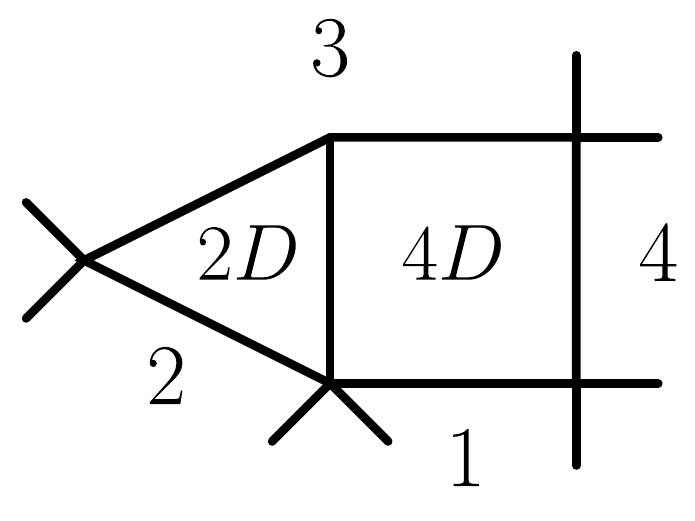}} \right] \nonumber\\ 
& 
\hspace{+3cm}
+  
\dlog \left(\frac{z \, \zb}{(1-z)(1-\zb)}\right) \ \   \raisebox{-28pt}{\includegraphics[scale=0.40]{figures/double_box_massive_2d4d.pdf}}\,. 
\hspace{-.5cm}
\end{align}
Alternatively, one can write the result more neatly in terms of $D=2$ boxes with (parity conjugate) chiral numerators
\begin{equation}
  \chi_\pm = (X_1X_2) (X_3 Y_L) - (X_1X_3) (X_2 Y_L) + (X_2X_3) (X_1 Y_L) \pm \langle Y_L X_1 X_2 X_3 \rangle\,,
\end{equation}
with the result
\begin{align}
   d I^{2\text{-loop}}_{4m}  & =  \dlog \left(\frac{z }{1-z}\right)  \  \raisebox{-28pt}{\includegraphics[scale=0.40]{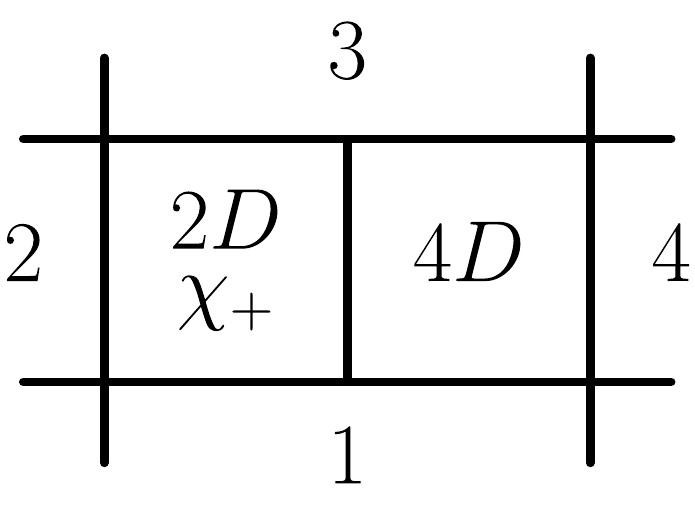}} -  
   \dlog \left(\frac{\zb}{1-\zb}\right)  \  \raisebox{-28pt}{\includegraphics[scale=0.40]{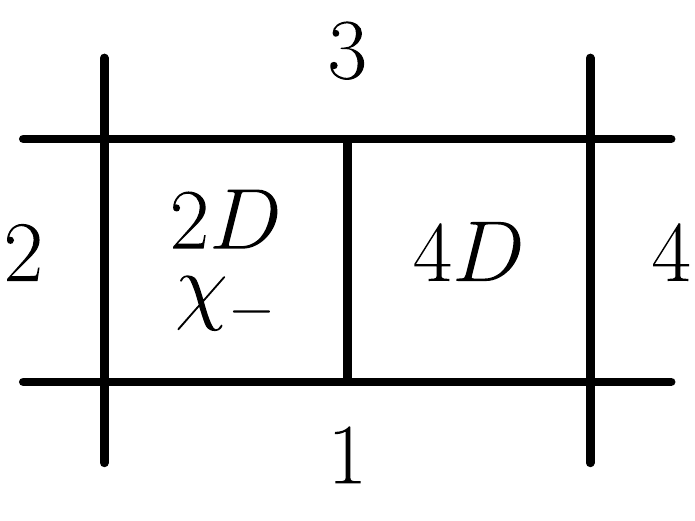}}\,.
\end{align}
%

%
\subsection{Higher-loop off-shell ladder}
\label{subsec:lloopladder}
%
Finally, with the differential equation for the two-loop ladder in hand, it is easy to derive differential equations for the general $L$-loop ladder integral\footnote{We thank Mark Spradlin for comments and suggestions.} in $D=4$
\begin{align}
I^{L\text{-loop}}_{4m} & = \raisebox{-28pt}{
\includegraphics[scale=.4]{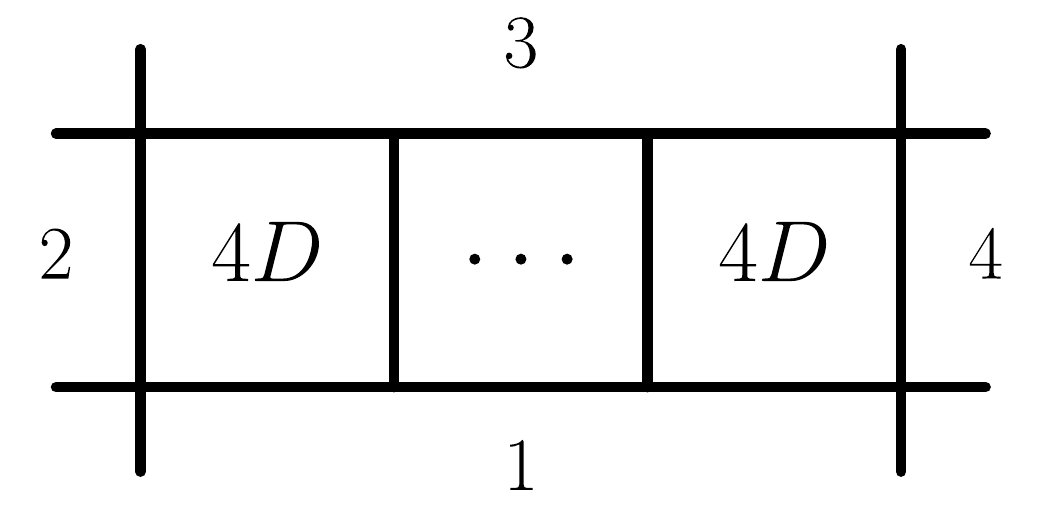}}\,.
\end{align}
It is not hard to check that the contribution analogous to $F_2$ in Eq.\eqref{eq:f1_def_2loop} also vanishes in this case. The partial fractioning of the localized contribution analogous to $F_1$ contribution is identical to the one in the previous section, so one obtains the differential equation
\vspace{-0.4cm}
\begin{align}
\hspace{-.5cm}
\begin{split}
   d I^{L\text{-loop}}_{4m}  & =  \dlog \left(\frac{z }{1-z}\right)  \  \raisebox{-28pt}{\includegraphics[scale=0.40]{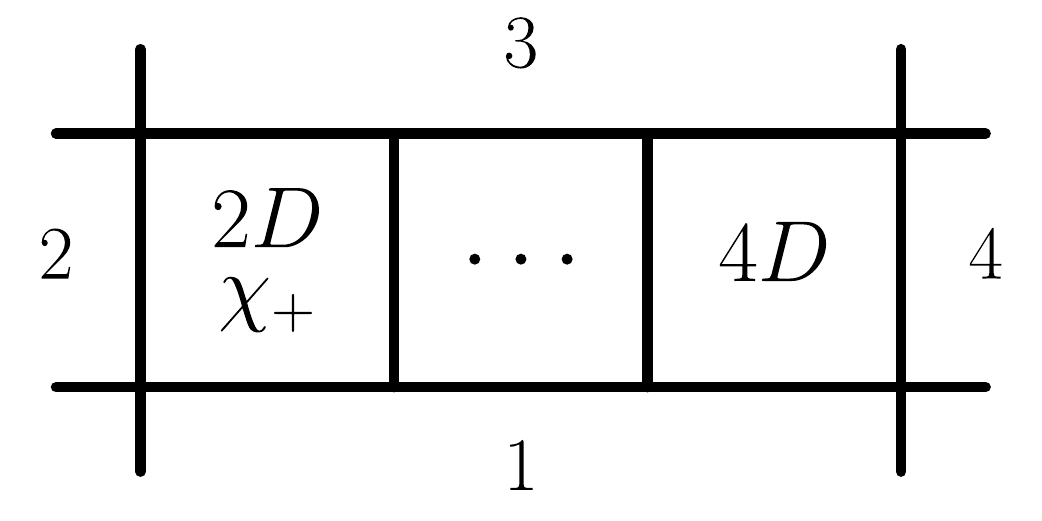}}  \\
   &-  
   \dlog \left(\frac{\zb}{1-\zb}\right)  \  \raisebox{-28pt}{\includegraphics[scale=0.40]{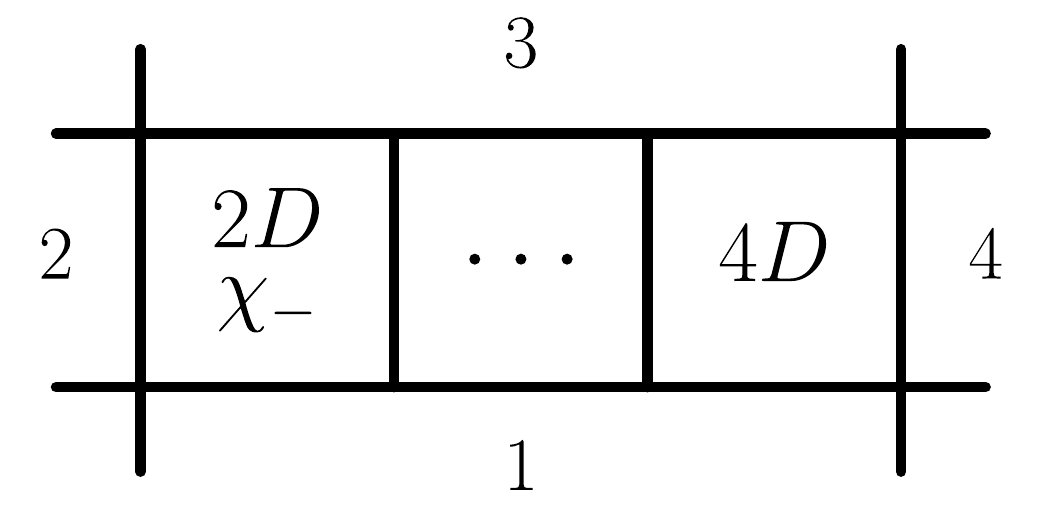}}\,.
\end{split}
\hspace{-.5cm}
\end{align}
It is well known that this integral satisfies a remarkable second-order differential equation which connects different loop orders \cite{Usyukina:1993ch,Broadhurst:2010ds,Drummond:2010cz,Basso:2017jwq,Caron-Huot:2018dsv}. Our differential equation, being first-order, identifies the intermediate weight-$(2L-1)$ objects in the symbol as mixed-dimensional chiral ladders. It would be interesting to study the differential equations of the mixed-dimensional integrals with our method and understand how they reproduce the know second order equation when combined with the ones derived above. It also would be natural to derive an analogous differential equation for the ``penta-ladder'' integrals, for which there is a similar understanding in terms of second-order equations \cite{Drummond:2010cz,Caron-Huot:2018dsv}. We leave this for future work.

\section{Conclusions}
\label{sec:conclusions}
\vspace{-.3cm}
%

In this work, we studied a novel use of the representation of Feynman integrals in terms of $\dlog$ forms. So far, $\dlog$ forms have played a major role in the context of $\N=4$ super Yang-Mills theory and (at the \emph{integrand} level) are crucial for modern ideas of reformulating perturbative quantum field theory in terms of geometric objects such as Grassmannians and the Amplituhedron. From a practical integration point of view, it has been conjectured, and empirically proven in numerous concrete examples, that $\dlog$ integrands lead to simplified differential equations in canonical form. In both contexts, the $\dlog$ forms were either not integrated or only used as identifying tool for convenient bases of master integrals. However, so far, no attempt has been made to utilize this novel representation of Feynman integrals directly for the integration process. In general, evaluating Feynman integrals is hard, and despite significant progress over the last few decades, we are still limited in both the low loop order as well as the number of kinematic scales involved in a given problem. Any new tool that could ultimately aid in understanding and evaluating Feynman integrals is therefore highly desirable. The expectation that $\dlog$ forms are ideally suited for integration purposes is related to the fact that they are extremely close to primitives already.

In this work, we make initial progress in evaluating $\dlog$ representations of Feynman integrals directly in loop-momentum space (or embedding space). In particular, we re-derive particularly simple differential equations for a special class of $D$-gon integrals in $D$ spacetime dimensions. These differential equations also appeared in the mathematical structure of volumes of hyperbolic simplices as early as in the 19$^{\text{th}}$ century. In modern form, the same differential equations also appeared in the context of motives due to a formula by Goncharov \cite{Goncharov:1996tate} and have been exported to physics by the works of Spradlin and Volovich \cite{Spradlin:2011wp} and later by others \cite{Arkani-Hamed:2017ahv,Abreu:2017enx}. 

Beyond the motivic one-loop differential equations, we were able to extend the applicability of the $\dlog$ algorithm to two loops and discussed several new features on a concrete example of the two-loop off-shell double box integral. Even though all the one-loop integrals as well as the two-loop example have been known for a long time, going forward, our analysis teaches a number of concrete lessons. First, we found that the differential equations are closely related to the canonical $\eps$ form, but the terms that naturally appear in the $\dlog$ context are identified with Feynman integrals in different spacetime dimensions. We found that higher-loop Feynman integrals with mixed dimension are interesting objects that should play a role in studying differential equations more generally. In some respect, to experts, this might not come as a too big of a surprise, as these mixed dimension integrals are suitable objects to manifest the transcendentality properties of the integrals. One further key aspect of the $\dlog$ differential equations derived in our work is that we never had to solve any large systems of integration-by-parts relations and all 
operations amounted to simple residue computations in order to extract the differential information of an integral. 

Along the way, we found a geometric meaning of the symbol entries of the Feynman integrals in terms of their cut geometry. We furthermore used residue theorems familiar from integrand considerations to reduce the possible final entries of an integral to a minimal set. 

Despite all our improved understanding of $\dlog$ forms, there are still a number of open problems that have to be solved in the future, before the $\dlog$ differential equations can be truly industrialized. First and foremost, up until now, finding compact $\dlog$ forms for a given integral that is expected to have only logarithmic singularities is still more an art than a science. Since this step is purely an integrand-level statement, further progress seems not out of reach. 

In this work, we only studied finite integrals (both in the infrared and the ultraviolet). It would be very interesting to get a handle at divergent integrals as well. Somewhat related, we only studied integrals, at order $\O(\eps^0)$, i.e. in integer dimensions. An extension of our procedure to include dimensional regularization in a straightforward manner would be highly desirable. It would then be interesting to understand the relation between the $\dlog$ differential equations derived in this paper and the differential equation in canonical form, and the recent applications of intersection theory to Feynman integrals \cite{Mizera:2017rqa,Mastrolia:2018uzb,Frellesvig:2019kgj,Frellesvig:2019uqt}. 

One additional question that naturally arises is, how the $\dlog$ story extends beyond the realm of generalized polylogarithms. It is by now well known that even in the simplest supersymmetric quantum field theories, the space of generalized polylogarithms is insufficient to describe scattering amplitudes and more complicated functions, such as elliptic integrals or integrals over Calabi-Yau manifolds appear. It would be extremely interesting to understand the elliptic and higher-complexity analogs of $\dlog$ integrands and what it would imply for certain `purity' or `transcendentality' statements of the more complicated function spaces. 

\medskip
\textbf{Note:} Some of the results discussed in this paper, have been known for some time to Simon Caron-Huot. 
The authors are indebted to Simon for his encouragement, generosity and help during different stages of this project.
\newpage

\vspace{-.5cm}
\section*{Acknowledgements}
\vspace{-.3cm}

We are grateful to Nima Arkani-Hamed, Zvi Bern, Ruth Britto, Simon Caron-Huot, Lance Dixon, 
Claude Duhr, Falko Dulat, Johannes Henn, Sebastian Mizera, Andrej Pokraka, Marcus Spradlin, Jaroslav Trnka, 
Anastasia Volovich, Ellis Ye Yuan, and Mao Zeng for enlightening discussions and to Zvi Bern, Lance Dixon, and Marcus Spradlin for comments on the manuscript. 
E.H.\ is grateful to the Mani L. Bhaumik Institute for Theoretical Physics at UCLA, for hospitality during various stages of this project, 
and the Aspen Center for Physics, which is supported by National Science Foundation grant PHY-1607611. 
The work of E.H.\ is supported by the U.S. Department of Energy (DOE) under contract DE-AC02-76SF00515.
J.P.-M.\ thanks the Mani L. Bhaumik Institute for generous support and SLAC for hospitality.
J.P.-M.\ is supported by the U.S.  Department of State through a Fulbright Scholarship. 
The authors also acknowledge support from the CERN theory group, and a stimulating environment that helped to finally bring this project to completion.

\appendix
\section{Integrals in embedding space on and off the null-cone}
\label{app:embedding}

In this appendix we summarize our conventions for the embedding space formalism \cite{Dirac:1936fq,Weinberg:2010fx}, following mostly \cite{Caron-Huot:2014lda} (see also \cite{Abreu:2017ptx}). We review how Feynman integrals in $D$ dimensions are represented in an $D+2$-dimensional embedding space, and how considering them off the null-cone (see below) naturally suggests a $\dlog$ form.

In the embedding space formalism, one considers $D+2$-dimensional projective space with homogeneous coordinates 
\begin{align}
\label{eq:embedding_coords}
X^M =\begin{pmatrix} X^\mu, & X^-, & X^+ \end{pmatrix}, 		
\end{align}
projectively identified $X^M \sim \lambda X^M$ with $\lambda \in \mathbb{C}^*$; and metric 
\begin{align}
\label{eq:embedding_metric}
(XY) = \eta_{MN}X^M Y^M = 2 X^\mu Y_\mu + X^+Y^-+X^-Y^+\,.
\end{align}
When there is no risk of confusion we will drop the round brackets that denote this inner product.
Compactified $D$-dimensional Minkowski space is identified with the quadric $(XX)=0$, also known as the projective null cone, $\Sigma_D$. In short, the embedding space formalism describes the correspondence between projective null vectors in (D+2) dimensions and points in compactified D-dimensional spacetime. 

We can define the metric on Minkowski space by choosing a special point, $I$, usually known as the point at infinity, and writing
\begin{align}
  \frac{(XY)}{(XI)(YI)} =  -(x-y)^2\,.
\end{align}
In a projectively invariant quantity only $(XY)$ remains and the propagators are effectively linearized. Note that apart from this definition \emph{poles at infinity} are treated in a uniform way with any other poles due to the conformal compactification (see e.g. \cite{Abreu:2017ptx}). A convenient gauge/coordinate choice is given by $(XI)=1$. Thus, fixing
\begin{align}
  I \equiv \begin{pmatrix} 0^\mu, & 1, & 0 \end{pmatrix}\,,
\end{align}
to each point $x^\mu$ in Minkowski space one associates a $D+2$ dimensional vector 
\begin{align}
 X^M = \begin{pmatrix} x^\mu, & -x^2, & 1 \end{pmatrix}\,.
\end{align}
For more general applications we are interested in cases where internal propagators are massive. Doing so requires relaxing the null condition and adding a mass to the $X^-$ component, such that $(XX) = 2 m^2$.

Loop integration variables can be treated in a similar fashion by introducing null vectors $Y_i$ in embedding space. Choosing the gauge above we have
\begin{align}
  \ell^\mu_i\mapsto Y_i = \begin{pmatrix}\ell^\mu_i, & -\ell^2_i, & 1 \end{pmatrix}\,.
\end{align}
The projectively invariant measure in embedding space can be written as
\begin{align}
\label{eq:embedding_unfixed_measure}
\frac{d^{D+2}Y }{\text{vol} \, \text{GL}(1)} =  \langle Y dY^{D+1}\rangle\,,
\end{align}
where 
\begin{equation}
  \langle Y dY^{D+1}\rangle = \frac{1}{(D+1)!} \, \epsilon_{I_1 I_2 \cdots I_{D+2}} \, Y^{I_1}\, dY^{I_2}\wedge \cdots \wedge dY^{I_{D+2}}
  \label{eq:holomorphic_measure}
\end{equation}
and $\text{vol}\, \text{GL}(1)$ reminds us that we have not yet gauge fixed the $GL(1)$ rescaling symmetry of $Y$. Leaving this gauge freedom unfixed turns out to be useful in certain computations when one does not want to commit to a certain coordinate chart. 

Integrals over Minkowski space just correspond to integrating over the $(YY)=0$ null cone
\begin{align}
\label{eq:embedding_measure}
\int\limits_{\Sigma_D} = \int \delta\left(\frac{1}{2}(YY)\right)
\end{align}
In the gauge above it is straightforward to relate the embedding space integration measure to the usual loop-momentum measure,
\begin{align}
\label{eq:loop_measure}
 \frac{ \langle Y dY^{D+1}\rangle}{(YI)^D} \delta\left(\frac{1}{2}(YY)\right) =  \frac{d^D\ell}{i\pi^{D/2}}
\end{align}
where we implicitly normalized by $i\pi^{D/2}$ when writing embedding space integrals.

Instead of using a delta function we can write an integral over the null-cone as a residue integral\footnote{We slightly abusing notation here, since here $\delta$ is not a function as in Eq.\eqref{eq:loop_measure} but a one-form.}
\begin{align}
\label{eq:loop_int}
\int \delta\left(\frac{1}{2}(YY)\right) \omega = \frac{1}{i\pi} \oint \dlog{(YY)}\wedge\omega
\end{align}
where the contour just calculates the residue at $(YY)=0$. Note that there is an ambiguity when writing the integrals in such a way, namely, a choice of how to extend the integrand, $\omega$, outside of the null-cone. The only requirement is that it gives the correct residue. In other words any two choices $\omega$ and $\omega'$ must have the same restriction to $\Sigma_D$
\begin{equation}
  \text{Res}{}_{\Sigma_D}[\omega] = \omega|_{\Sigma_D} = \omega'|_{\Sigma_D} = \text{Res}{}_{\Sigma_D} [\omega']
\end{equation}
This freedom will turn out to be very useful. As explained below, it will demystify the appearance of special propagators in the $\dlog$ forms of loop integrals.

\subsection*{A detailed example: The four-dimensional off-shell box}
\label{app:example}

As an example we will describe the different ways of writing the four dimensional off-shell box integral in embedding space. In dual variables \cite{Drummond:2006rz} this integral is given by
\begin{equation}
  I_4 =  \raisebox{-41pt}{\includegraphics[width=0.2\textwidth]{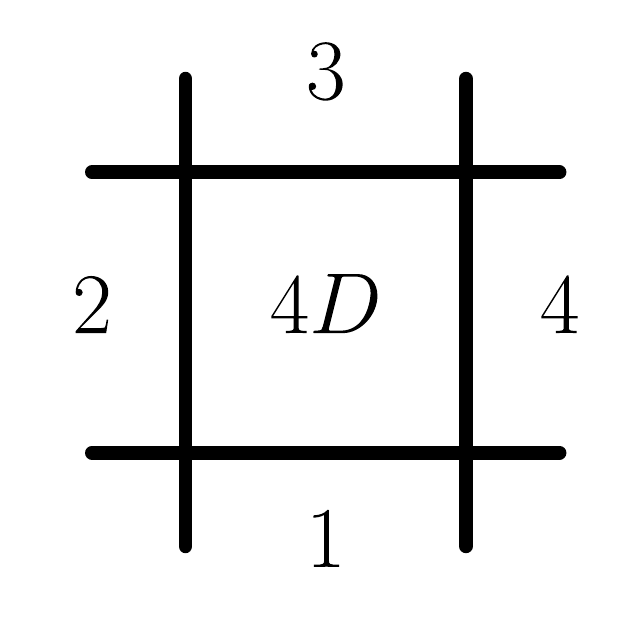}}=  \int \frac{d^4y}{i\pi^{2}} \frac{\sqrt{\text{det}(x_i-x_j)^2}}{(y-x_1)^2 (y-x_2)^2 (y-x_3)^2 (y-x_4)^2}
\end{equation}
The integrand can easily be transcribed to embedding space using the prescription above,
\begin{equation}
  I_4 = \int\limits_{\Sigma_4} \frac{ \langle Y d^5Y\rangle \sqrt{-\text{det}\,Q}}{(YX_1)(YX_2)(YX_3)(YX_4)}\,,
\end{equation}
where we introduced the Gram matrix $Q_{ab} = (X_aX_b)$ with $a,b=1,\ldots,4$. We see that in $D=4$ the dependence on the point at infinity $(YI)$ cancels with the factor in the measure \eqref{eq:loop_measure}. For triangle integrals this is not the case and an additional pole $(YI)$ remains. In the common gauge choice $(YI)=1$, this pole is hidden but it is this term that is often referred to as the \emph{pole at infinity} for triangle integrals, see e.g. \cite{Forde:2007mi,Bern:2014kca}. 

We normalized the box so that it has unit leading singularities and a $\dlog$ form. This has the side effect of making it parity odd, since it changes sign if we flip the sign of the square root. This can be made more manifest in embedding space as follows. Consider the extended Gram matrix $Q^*_{ab} = (X_aX_b)$ with $a,b=1,\ldots,4,+,-$ where $X_\pm$ are the two solutions to the maximal cut of the box. Since $(X_\pm X_i)=0$ this matrix is block diagonal and its determinant factorizes  $\det Q^* = - (X_+X_-)^2 \det Q$. Note also that we can write the determinant of  $Q^*$ in terms of the skew-symmetric tensor
\begin{equation}
  \det Q^* = \langle X_1 X_2 X_3 X_4 X_+ X_- \rangle^2
\end{equation}
So we can rewrite the box integral in embedding space as
\begin{equation}
 I_4= \int \frac{\langle Y d^{5}Y\rangle \;\langle X_1 X_2 X_3 X_4 X_- X_+ \rangle}{(X_+X_-)(YX_1)(YX_2)(YX_3)(YX_4)} 
\end{equation}
Parity exchanges $X_+ \leftrightarrow X_-$, which manifestly flips the sign of the numerator.

Let us now go off the null cone and rewrite the box as a residue integral. 
\begin{equation}
  I_4= \frac{1}{i\pi} \oint \frac{\langle Y d^{5}Y\rangle \; \langle X_1 X_2 X_3 X_4 X_+ X_- \rangle}{(YY)(X_+X_-)(YX_1)(YX_2)(YX_3)(YX_4)} =  \frac{1}{2\pi i}\oint \omega_{\rm can} \,.
\end{equation}
We have chosen to extend the integrand away from the null cone in a canonical way, that is, without adding any extra pieces. This form has new poles away from the original contour. To explicitly show how this comes about, it is best to use the completeness relation in the embedding space coordinates,
\begin{align}
\eta^{MN} = \sum\limits^{6}_{a,b=1} c_{ab} X^{M}_a X^{N}_b
\end{align}
expanded in a basis of $6$ vectors in embedding space. We choose as the basis the four dual points, $X_i$, and $X_\pm$.  It is easy to see that with this choice
\begin{equation}
  c_{ab} = \left\{ 
  \begin{array}{cl}
    Q^{-1}_{ab} & \text{for} \quad a,b=1,\ldots,4 \\
    (X_+ X_-)^{-1} & \text{for} \quad (a,b)= (+,-) \hspace{5pt} \text{or} \hspace{5pt}  (a,b)= (-,+)\\
    0   & \text{otherwise}
  \end{array} \right.
\end{equation}
Using this completeness relation the extra pole can be written as
\begin{equation}
  (YY)= 2 \frac{(YX_+)(Y X_-)}{(X_+X_-)} + \sum\limits^{D}_{a,b=1} Q^{-1}_{ab} (YX_a)(YX_b)
\end{equation}
Now it is easy to see that the residue of the form above on the maximal cut $(YX_i) =0$ is
\begin{equation}
  \text{Res}{}_C [\omega_{\rm can}] = \dlog\left(\frac{YX_+}{Y X_-}\right)
\end{equation}
which still has poles when $(YX_+)=0$ or $(Y X_-)=0$. With this in mind one can show that
\begin{align}
  \label{eq:dlogcan}
\begin{split}
\hspace{-.1cm}
  \omega_{\text{can}} &= \dlog\left(\frac{YX_1}{Y X_\pm}\right)\dlog\left(\frac{YX_2}{Y X_\pm}\right)\dlog\left(\frac{YX_3}{Y X_\pm}\right)\dlog\left(\frac{YX_4}{Y X_\pm}\right)\dlog\frac{YY}{(Y X_\pm)^2} 
 \end{split}
\end{align}
The arguments of the $\dlog$s seemingly obscure the conformal invariance. This can be easily remedied, by introducing additional arbitrary points in the arguments. For instance
\begin{equation}
  \frac{YX_1}{Y X_+} \rightarrow \frac{(YX_1)(X_+ Z)}{(Y X_+)(X_1 Z)}
\end{equation}
The choice in \eqref{eq:dlogcan} corresponds to choosing $Z=I$, the point at infinity, since in our gauge $(X_i I) = (X_\pm I) =1$. Unless necessary, we will not write these extra points explicitly.

There are other ways to continue the integrand off the null-cone, which correspond to the different $\dlog$ forms that one usually writes in momentum space. For instance one can write
\begin{equation}
  \omega_+ = \dlog(YY)\,\dlog\left(\frac{YX_1}{Y X_+}\right)\dlog\left(\frac{YX_2}{Y X_+}\right)\dlog\left(\frac{YX_3}{Y X_+}\right)\dlog\left(\frac{YX_4}{Y X_+}\right)
\end{equation}
which is related to $\omega_{\text{can}}$ as follows
\begin{equation}
  \omega_+-\omega_{\text{can}} =  \dlog(YX_+)  \dlog(YX_1)\dlog(YX_2)\dlog(YX_3)\dlog(YX_4)
\end{equation}
This is just $\omega_{\text{can}}$ with the pole at $(YX_+)$ subtracted. Similarly one can construct a form $\bar\omega_{+}$ by subtracting the other pole, or 
\begin{equation}
  \omega_{+/-} = \frac12 \dlog(YY)\,  \dlog\left(\frac{YX_1}{Y X_2}\right)\dlog\left(\frac{YX_2}{Y X_3}\right)\dlog\left(\frac{YX_3}{Y X_4}\right)\dlog\left(\frac{YX_{+}}{Y X_-}\right)
\end{equation}
which cancels both poles. Note that $\omega_{+/-} =-(\omega_+ - \bar\omega_{+})/2$. Finally, an interesting choice is the following
\begin{equation}
\omega_{\text{can}}' = \dlog\left(\frac{(YY)(X_+ X_-)}{(YX_{+})(Y X_-)}\right)  \dlog\left(\frac{YX_1}{Y X_+}\right)\dlog\left(\frac{YX_2}{Y X_+}\right)\dlog\left(\frac{YX_3}{Y X_+}\right)\dlog\left(\frac{YX_4}{Y X_+}\right)
\end{equation}
which satisfies
\begin{align}
  \omega_{\text{can}}' - \omega_{\text{can}} &= \dlog\left(\frac{YX_1}{Y X_+}\right)\dlog\left(\frac{YX_2}{Y X_+}\right)\dlog\left(\frac{YX_3}{Y X_+}\right)\dlog\left(\frac{YX_4}{Y X_+}\right)\dlog\left(\frac{YX_-}{Y X_+}\right) \nonumber \\
  & = \frac12 \frac{\langle Y d^{5}Y\rangle \; \langle X_1 X_2 X_3 X_4 X_+ X_- \rangle }{(YX_+)(YX_-)(YX_1)(YX_2)(YX_3)(YX_4)}
\end{align}
or equivalently
\begin{equation}
  \omega_{\text{can}}' = \omega_{\text{can}} \left(1 - \frac12 \frac{(YY)(X_+ X_-)}{(YX_+)(Y X_-)} \right)
\end{equation}
Note that this form does not have a residue on the maximal cut, but has residues at the hyperplanes where $(YY)$, $(YX_+)$ or $(YX_-)$ vanish. The difference between any two of these forms vanishes upon taking the residue at $(YY)=0$, so choosing one or another is a matter of convenience. Making the right choice can greatly simplify the task of partial fractioning a given rational form, an operation which plays an important role in the main text.

\bibliographystyle{JHEP}
\phantomsection         
            \bibliography{amplitude_refs.bib}
            \clearpage

\end{document}